\documentclass[
]{aa}

\usepackage{graphicx}
\graphicspath{{PLOT/}}

\usepackage{txfonts}
\usepackage{ctable}
\usepackage{natbib}

\usepackage[plainpages=False, bookmarks=False, colorlinks=True,
citecolor=blue, linkcolor=blue, draft=True]{hyperref}

\newcommand{\Msol}{\ensuremath{\mathrm{M}_\odot}}

\newcommand{\Ha}{\ensuremath{\text{H}\alpha}}

\newcommand{\SigmaHa}{\ensuremath{\Sigma_{\Ha}}}

\newcommand{\sS}{\ensuremath{\mathrm{Ia}\alpha}}
\newcommand{\sN}{\ensuremath{\mathrm{Ia}\epsilon}}

\newcommand{\hostmass}{\ensuremath{\log(\mathrm{M}/\Msol)}}

\newcommand{\DcHRMS}{\ensuremath{
    \delta \langle M_B^{\mathrm{corr}}\rangle(H-L)}}
\newcommand{\Mone}{\ensuremath{M_{1}}}
\newcommand{\Mtwo}{\ensuremath{M_{2}}}
\newcommand{\cHR}{\ensuremath{\Delta M_B^{\mathrm{corr}}}}
\newcommand{\cH}{\ensuremath{M_B^{\mathrm{corr}}}}
\newcommand{\DcHR}{%
  \ensuremath{\delta \langle M_B^{\mathrm{corr}}\rangle(\sN-\sS) }}

\begin{document}

\title{Evidence of Environmental Dependencies of Type Ia
    Supernovae from the Nearby Supernova Factory indicated by Local
    \Ha{}}

\author
{
  {M.~Rigault}\inst{\ref{ipnl}}
  \and {Y.~Copin}\inst{\ref{ipnl}}
  \and {G.~Aldering}\inst{\ref{lbnl}}
  \and {P.~Antilogus}\inst{\ref{lpnhe}}
  \and {C.~Aragon}\inst{\ref{lbnl}}
  \and {S.~Bailey} \inst{\ref{lbnl}}
  \and {C.~Baltay}\inst{\ref{yale}}
  \and {S.~Bongard}\inst{\ref{lpnhe}} 
  \and {C.~Buton}\inst{\ref{bonn}}
  \and {A.~Canto}\inst{\ref{lpnhe}}
  \and {F.~Cellier-Holzem}\inst{\ref{lpnhe}}
  \and {M.~Childress}\inst{\ref{austra}}
  \and {N.~Chotard}\inst{\ref{ipnl}}
  \and {H.~K. Fakhouri}\inst{\ref{lbnl},\ref{berkphys}}
  \and {U.~Feindt\inst{\ref{bonn}}}
  \and {M.~Fleury}\inst{\ref{lpnhe}}
  \and \\ {E.~Gangler}\inst{\ref{ipnl}}
  \and {P.~Greskovic}\inst{\ref{bonn}} 
  \and {J.~Guy}\inst{\ref{lpnhe}}
  \and  {A.~G.~Kim}\inst{\ref{lbnl}}
  \and {M.~Kowalski}\inst{\ref{bonn}}
  \and {S.~Lombardo}\inst{\ref{bonn}} 
  \and {J.~Nordin}\inst{\ref{lbnl},\ref{berk_ssl}}
  \and {P.~Nugent}\inst{\ref{berkccc},\ref{berk_astro}}
  \and \\{R.~Pain}\inst{\ref{lpnhe}}
  \and {E.~P\'econtal}\inst{\ref{cral}}
  \and {R.~Pereira}\inst{\ref{ipnl}}
  \and {S.~Perlmutter}\inst{\ref{lbnl},\ref{berkphys}}
  \and {D.~Rabinowitz}\inst{\ref{yale}}
  \and {K.~Runge}\inst{\ref{lbnl}}
  \and {C.~Saunders}\inst{\ref{lbnl}}
  \and \\{R.~Scalzo} \inst{\ref{austra}}
  \and {G.~Smadja}\inst{\ref{ipnl}}
  \and {C.~Tao}\inst{\ref{tsinghua},\ref{cppm}}
  \and  {R.~C.~Thomas}\inst{\ref{berkccc}}
  \and {B.~A.~Weaver}\inst{\ref{nyu}}
  \\(The Nearby Supernova Factory)
}

\institute{
  Universit\'e de Lyon, 69622, France; Universit\'e de Lyon 1, France;
  CNRS/IN2P3, Institut de Physique Nucl\'eaire de Lyon,
  France
  \email{rigault@ipnl.in2p3.fr}\label{ipnl}
  \and
  Physics Division, Lawrence Berkeley National Laboratory,
  1 Cyclotron Road, Berkeley, CA 94720, USA\label{lbnl}
  \and
  Laboratoire de Physique Nucl\'eaire et des Hautes \'Energies,
  Universit\'e Pierre et Marie Curie Paris 6, Universit\'e Paris
  Diderot Paris 7, CNRS-IN2P3, 4 place Jussieu, 75252 Paris Cedex 05,
  France\label{lpnhe}
  \and
  Department of Physics, Yale University,
  New Haven, CT 06520-8121, USA\label{yale}
  \and
  Physikalisches Institut, Universit\"at Bonn,
  Nu\ss allee 12, 53115 Bonn, Germany\label{bonn}
  \and
    Research School of Astronomy and Astrophysics,
    The Australian National University,
    Mount Stromlo Observatory,
    Cotter Road, Weston Creek ACT 2611 Australia\label{austra}
  \and
  %
  %
  %
  Department of Physics, University of California Berkeley,
  366 LeConte Hall MC 7300, Berkeley, CA, 94720-7300, USA\label{berkphys}
  \and
  Space Sciences Laboratory, University of California Berkeley,
  7 Gauss Way, Berkeley, CA 94720, USA\label{berk_ssl}
  \and
  Computational Cosmology Center, Computational Research Division,
  Lawrence Berkeley National Laboratory, 1 Cyclotron Road MS~50B-4206,
  Berkeley, CA, 94611, USA\label{berkccc}
  \and
  Department of Astronomy, University of California Berkeley,
  B-20 Hearst Field Annex \# 3411, Berkeley, CA, 94720-34110 
  \label{berk_astro} 
  \and
  Centre de Recherche Astronomique de Lyon, Universit\'e Lyon 1,
  9 Avenue Charles Andr\'e, 69561 Saint Genis Laval Cedex, France\label{cral}
  \and
  Tsinghua Center for Astrophysics, Tsinghua University, Beijing
  100084, China\label{tsinghua}
  \and
  Centre de Physique des Particules de Marseille, 163, avenue de
  Luminy - Case 902 - 13288 Marseille Cedex 09, France\label{cppm}
  \and
  Center for Cosmology and Particle Physics,
    New York University,
    4 Washington Place, New York, NY 10003, USA\label{nyu}
}


\abstract{
  Use of Type Ia supernovae (SNe~Ia) as distance indicators has proven
  to be a powerful technique for measuring the dark energy equation of
  state.  However, recent studies have highlighted potential biases
  correlated with the global properties of their host galaxies,
  sufficient in size to induce systematic errors into such
  cosmological measurements if not properly treated.}{
  We study the host galaxy regions in close proximity to SNe~Ia in
  order to analyze relations between the properties of SN~Ia events
  and environments most similar to where their progenitors formed.  In
  this paper we will focus on local \Ha{} emission as an indicator of
  young progenitor environments.}{
  The Nearby Supernova Factory has obtained flux-calibrated spectral
  timeseries for SNe~Ia using integral field spectroscopy.  These
  observations enable the simultaneous measurement of the SN and its
  immediate vicinity.  For 89~SNe~Ia we measure or set limits on \Ha{}
  emission tracing ongoing star formation within a 1~kpc radius around
  each SN.  This constitutes the first direct study of the local
  environment for a large sample of SNe~Ia also having accurate
  luminosity, color and stretch measurements.}{
  Our local star formation measurements provide several critical new
  insights. 
  We find that SNe Ia with local \Ha{} emission are redder
  by $0.036\pm0.017$ mag, and that the previously-noted correlation
  between stretch and host mass is entirely driven by the SNe Ia
  coming from locally passive environments, in particular at the
  low-stretch end. There is no such trend for SNe Ia having locally
  star-forming environments. Our most important finding is that the
  mean standardized brightness for SNe Ia with local \Ha{} emission is
  $0.094\pm0.031$ mag fainter on average than for those without.  This
  offset arises from a bimodal structure in the Hubble residuals, with
  one mode being shared by SNe Ia in all environments and the other
  one being exclusive to SNe Ia in locally passive environments. This
  structure also explains the previously-known host-mass bias. We
  combine the star-formation dependence of this bimodality with the
  cosmic star-formation rate to predict changes with redshift in the
  mean SN Ia brightness and the host-mass bias. The strong change
  predicted is confirmed using high-redshift SNe Ia from the
  literature.  }{
  The environmental dependences in SN~Ia Hubble residuals and color
  found here point to remaining systematic errors in the
  standardization of SNe~Ia. In particular, the observed brightness
  offset associated with local \Ha{} emission is predicted to cause a
  significant bias in current measurements of the dark energy equation
  of state.  Recognition of these effects offers new opportunities to
  improve SNe~Ia as cosmological probes. For instance, we note that
  SNe~Ia associated with local \Ha{} emission are more homogeneous,
  having a brightness dispersion of only $0.105 \pm 0.012$~mag.
}

\titlerunning{Local H$\alpha$ analysis of Type Ia supernovae}
\authorrunning{M.~Rigault \& the Nearby Supernova Factory}

\maketitle

\section{Introduction}

Luminosity distances from Type~Ia supernovae (SNe~Ia) were key to the
discovery of the accelerating expansion of the universe
\citep{perlmutter_measurements_1999, riess_observational_1998}.  Among
the current generation of surveys more than 600 spectroscopically
confirmed SNe~Ia are available for cosmological analyses
\citep[e.g.][]{suzuki_hubble_2012}.  Thus, even today SNe~Ia remain
the strongest demonstrated technique for measuring the dark energy
equation of state.

The fundamental principle behind the use of these standardized candles
is that the standardization does not change with redshift.  SNe~Ia
have an observed $M_{B}$ dispersion of approximately 0.4~mag, which
makes them naturally good distance indicators.  Empirical light-curve
fitters such as SALT2 \citep{guy_salt2:_2007, guy_supernova_2010} or
MLCS2K2 \citep{jha_improved_2007} correct $M_{B}$ for the
``brighter-slower'' and ``brighter-bluer'' relation
\citep{phillips_absolute_1993, riess_precise_1996,
  tripp_two-parameter_1998}.  This stretch (or $x_{1}$) and color
($c$) standardization enables the reduction of their magnitude
dispersion down to $\approx 0.15$~mag.

However, a major issue remains: despite decades of study, their
progenitors are as yet undetermined. \citep[See][for a detailed
review.]{maoz_type-ia_2011} Like all stars, it is expected that these
progenitors will have a distribution of ages and metal abundances, and
these distributions will change with redshift.  These factors in turn
may effect details of the explosion, leading to potential bias in the
cosmological measurements.  The remaining 0.15~mag ``intrinsic''
scatter in SN~Ia standardized brightnesses is a direct indicator that
hidden variables remain.  Host galaxy dust -- and peculiar velocities
if the host is too nearby -- complicate the picture.

Several studies have found that the distribution of SNe~Ia light curve
stretches differs across host galaxy total stellar mass
\citep{hamuy_search_2000, neill_local_2009, sullivan_dependence_2010}
and global specific star formation rate (sSFR)
\citep{lampeitl_effect_2010,
  konishi_dependences_2011}. \citet{lampeitl_effect_2010} concluded
that the distribution of SNe~Ia colors appears to be independent of
host star-forming properties, even though more dust is expected in
actively star-forming environments. Although, in
\citet{childress_host_2012} we found that SNe~Ia colors do correlate
with host metallicity; this may be intrinsic, but also metals are a
necessary ingredient for dust formation.  Stretch is correlated with
observed $M_{B}$, so after standardization the influence of this
environmental property disappears.

A dependence of corrected Hubble residuals on host mass is now
well-established \citep{kelly_hubble_2010, sullivan_dependence_2010,
  gupta_improved_2011, childress_host_2012, johansson_sne_2012}. This
has been modeled as either a linear trend or a sharp step in corrected
Hubble residuals between low- and high-mass hosts.  In
\citet{childress_host_2012} we established that a ``mass step'' at
$\hostmass=10.2$ gives a much better fit than a line, and we found
that the RMS width of the transition is only 0.5~dex in mass.  Because
the mass of a galaxy correlates with its metallicity, age, and sSFR
\citep[see][respectively]{tremonti_origin_2004, gallazzi_ages_2005,
  perez-gonzalez_stellar_2008}, this mass step is most likely driven
by an intrinsic SN progenitor variation.  For instance, a brightness
offset between globally star-forming and globally passive galaxies
provides a fair phenomenological description of the mass step
\citep{dandrea_spectroscopic_2011},
 being driven by the sharp change in the fraction of star-forming hosts at
$\hostmass\sim10$ present in the local universe
\citep{childress_host_2012}.

\begin{figure}
  \centering
  \includegraphics[width=\linewidth,clip]{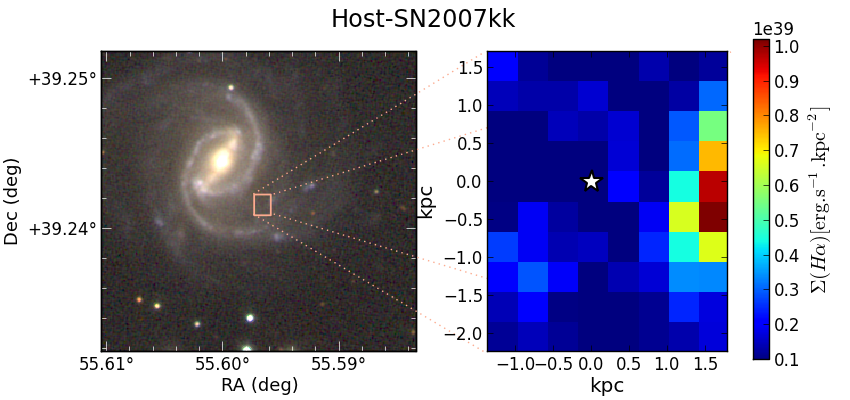}
  \includegraphics[width=\linewidth,clip]{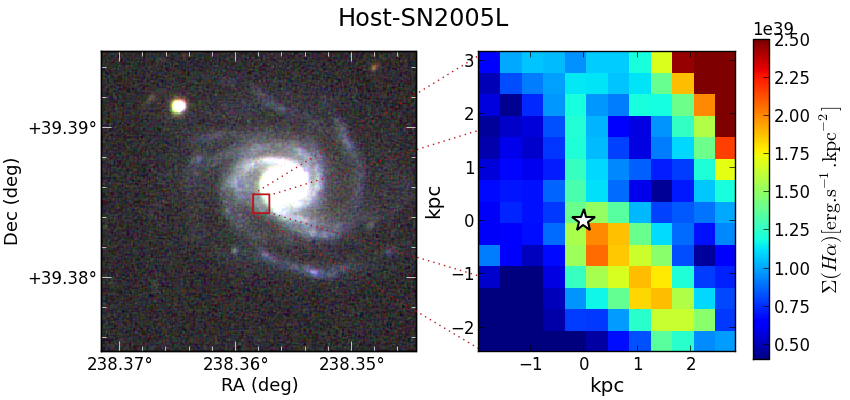}
  \caption{\emph{Top and Bottom:} The hosts of SN~2007kk (UGC~2828) and
    SN~2005L (MCG+07-33-005), respectively, both classified as
    globally star-forming \citep{childress_data_2013}.  \emph{Left:}
    color images made using observations from SNIFS and SDSS-III
    \citep{aihara_eighth_2011}.  On both images the field-of-view of
    SNIFS, centered on the SN position (white-star marker), is
    indicated by the red central square.  \emph{Right:} \Ha{} surface
    brightness maps of the SN vicinities (the generation of these maps
    is detailed in Sect.~\ref{sec:Map_creation}).  SN~2007kk occurred
    in a passive environment more than 1.5~kpc from the closest
    star-forming region, while SN~2005L is located at the edge of such
    a region.}
  \label{fig:global_vs_local}
\end{figure}

Though, by analyzing global properties of the host galaxy, the
aforementioned analyses are limited in the interpretation of their
results.  The measured quantities -- gas metallicity, star formation
rate, etc. -- are light-weighted.  Thus global analyses are most
representative of galaxy properties near the core, which can be
significantly different than the actual SN environment.  This is
illustrated in Fig.~\ref{fig:global_vs_local}: inside these two spiral
star-forming galaxies, a SN occurred either in an old passive
inter-arm environment (SN~2007kk) or inside a star-forming one
(SN~2005L).

In this work we analyze the host galaxy regions in the immediate
vicinity for a large sample of SNe~Ia from the Nearby Supernova
Factory \citep[SNfactory,][]{aldering_overview_2002}.  Our integral
field spectrograph accesses the local environment of observed SNe, and
therefore probes local host properties such as gas and stellar
metallicities and star formation history.  While
\cite{stanishev_type_2012} conducted such a study by looking at the
metallicity of the local environments of a sample of seven nearby
SNe~Ia, ours is the first such large-scale study.

Delay-time distribution studies \citep{scannapieco_type_2005,
  mannucci_supernova_2005, mannucci_two_2006, sullivan_rates_2006}
predict that a fraction of SNe~Ia, known as ``prompt'' SNe, should be
associated with young stellar populations.  The rest, referred to as
``tardy'' or ``delayed'' SNe, should be related to older stars.
Individual star-forming regions (\ion{H}{ii}~regions) have a typical
lifetime of a few~Myr \citep{alvarez_h_2006}, much smaller than
expected --~even for the fastest~-- SN~Ia progenitor systems
\citep[few tens of Myr,][]{girardi_evolutionary_2000}.  It is
therefore impossible to make a physical connection between a SN and
the \ion{H}{ii}~region in which its progenitor formed.  However such
star forming regions are gathered in groups \citep[see for instance
M~51 in][]{lee_h_2011}, concentrated in spiral arms whose lifetimes
are longer than the time scale for a prompt SN.  \citep[See][for
details on the star formation history of
galaxies.]{kauffmann_dependence_2003} This motivated us to employ
\Ha{} for our first analysis of SNe~Ia local environments. \Ha{} will
be taken to indicate active local star formation, and the likelihood
that a given progenitor was especially young.

Long before the ``prompt'' and ``tardy'' distinction, association with
\ion{H}{ii}~regions or spiral arms was a common approach for
understanding the stellar populations from which SNe formed.  For
instance, with small samples of SNe, \cite{bartunov_distribution_1994}
and later \cite{james_h_2006} showed that type~Ia SNe were less
associated with such regions than are core-collapse SNe.  A key
ingredient missing in such studies needed to understand SNe~Ia in
detail is accurate standardized luminosities, which we have for our
sample.  Also, in light of the ``prompt''/``tardy'' dichotomy, it will
be important to consider the possibility that SNe~Ia will fall into
discrete subsets based on the strength of star formation in their
local environment.

After the presentation of our data (Sect.~\ref{sec:Data}) and our
method for quantifying neighboring star-forming activity
(Sect.~\ref{sec:Analysis}), we investigate the correlation of SNe~Ia
light-curve parameters with the local environment
(Sect.~\ref{sec:Results}).  This includes a review of how our local
environment results compare with the aforementioned global host galaxy
studies.  In particular, we present evidence for excess dust
obscuration of SNe~Ia in actively star-forming environments.  We also
show that the corrected SN~Ia brightnesses depend significantly on the
local \Ha{} environment.  We discuss the implications of our results
in Sect.~\ref{sec:Discuss}, where the standardization of SNe~Ia is
examined.  We then compare our local \Ha{} analysis with previous
results concerning the total host stellar mass, finding important
clues to the origin of the mass step.  In
Sect.~\ref{sec:Cosmo_consequences}, we estimate the impact of the
local trends on the cosmological parameters using a simple model, and
use this model to show redshift evolution in SNe~Ia from the
literature.  We further demonstrate the existence of a SN~Ia subset
having a significantly reduced brightness dispersion.  Finally, in
Sect.~\ref{sec:Assumption_discussion} we test the robustness of our
results, and end with a summary in Sect.~\ref{sec:Conclusion}.

\section{Integral Field Spectroscopy of Immediate
  SN~Ia Neighborhoods}
\label{sec:Data}

\subsection{SNIFS and the SNfactory}
\label{sec:SNIFS}

The SNfactory has observed a large sample of nearby SNe
and their immediate surroundings using the SuperNova Integral Field
Spectrograph \citep[SNIFS,][]{lantz_snifs_2004} installed on the
UH~2.2m telescope (Mauna Kea).  SNIFS is a fully integrated instrument
optimized for semi-automated observations of point sources on a
structured background over an extended optical window at moderate
spectral resolution.  The integral field spectrograph (IFS) has a
fully-filled $6\farcs4 \times 6\farcs4$ spectroscopic field-of-view
subdivided into a grid of $15 \times 15$ contiguous square spatial
elements (spaxels).  The dual-channel spectrograph simultaneously
covers 3200--5200~\AA{} ($B$-channel) and 5100--10\,000~\AA{}
($R$-channel) with 2.8 and 3.2~\AA{} resolution, respectively.  The
data reduction of the $x,y,\lambda$ data cubes was summarized by
\cite{aldering_nearby_2006} and updated in Sect.~2.1 of
\cite{scalzo_nearby_2010}.  A preview of the flux calibration is
developed in Sect.~2.2 of \cite{pereira_spectrophotometric_2013},
based on the atmospheric extinction derived in
\cite{buton_atmospheric_2013}.

For every followed SN, the SNfactory creates a spectro-photometric
time series typically composed of $\sim$13~epochs, with the first
spectrum taken on average three days before maximum light in $B$
\citep{bailey_using_2009, chotard_reddening_2011}.  In addition,
observations are obtained at the SN location at least one year after
the explosion to serve as a final reference to enable the subtraction
of the underlying host.  For this analysis, we have gathered a
subsample of 119~SNe~Ia with good final references and properly
measured light-curve parameters, including quality cuts suggested by
\cite{guy_supernova_2010}.

\subsection{Local Host Observations}
\label{sec:split_component}

The SNfactory software pipeline provides flux calibrated
$x,y,\lambda$-cubes corrected for instrumental and atmospheric
responses \citep{buton_atmospheric_2013}.  Each spaxel of an SN~cube
includes three components, each one characterized by its own spatial
signature:
\begin{enumerate}
\item the SN is a pure point source, located close to the center of
  the SNIFS field of view (FoV);
\item the night sky spectrum is a spatially flat component over the
  full FoV;
\item the host galaxy is a (potentially) structured background.
\end{enumerate}

In this section, we present the algorithm used to disentangle the
three components.  In Sect.~\ref{sec:SN-subtraction}, we describe the
method used to subtract the SN component from the original cubes, and
in Sect.~\ref{sec:sky-subtraction} our sky subtraction procedure,
using a spectral model for the sky that prevents host-galaxy signal
contamination.  All cubes from the same host are finally combined to
produce high signal-to-noise (S/N) cubes and spectra; this process is
detailed in Sect.~\ref{sec:merge-cubes}.

\subsubsection{SN Subtraction}
\label{sec:SN-subtraction}

The point source extraction from IFS data requires the proper
subtraction of any structured background.  Using a 3D deconvolution
technique, \cite{bongard_3d_2011} show how we construct a
``seeing-free'' galaxy $x,y,\lambda$-cube from final reference
exposures taken after the SN~Ia has vanished (typically one year after
maximum light).  This model, after proper registration and
reconvolution with the appropriate seeing, is subtracted from each
observed cube to leave the pure point-source component plus a spatial
constant similar to a sky signal.  Three dimensional~PSF photometry is
then applied to the host-subtracted cube to extract the point source
spectrum \citep[e.g.][]{buton_atmospheric_2013}.

\begin{figure}
  \centering
  \includegraphics[width=\linewidth]{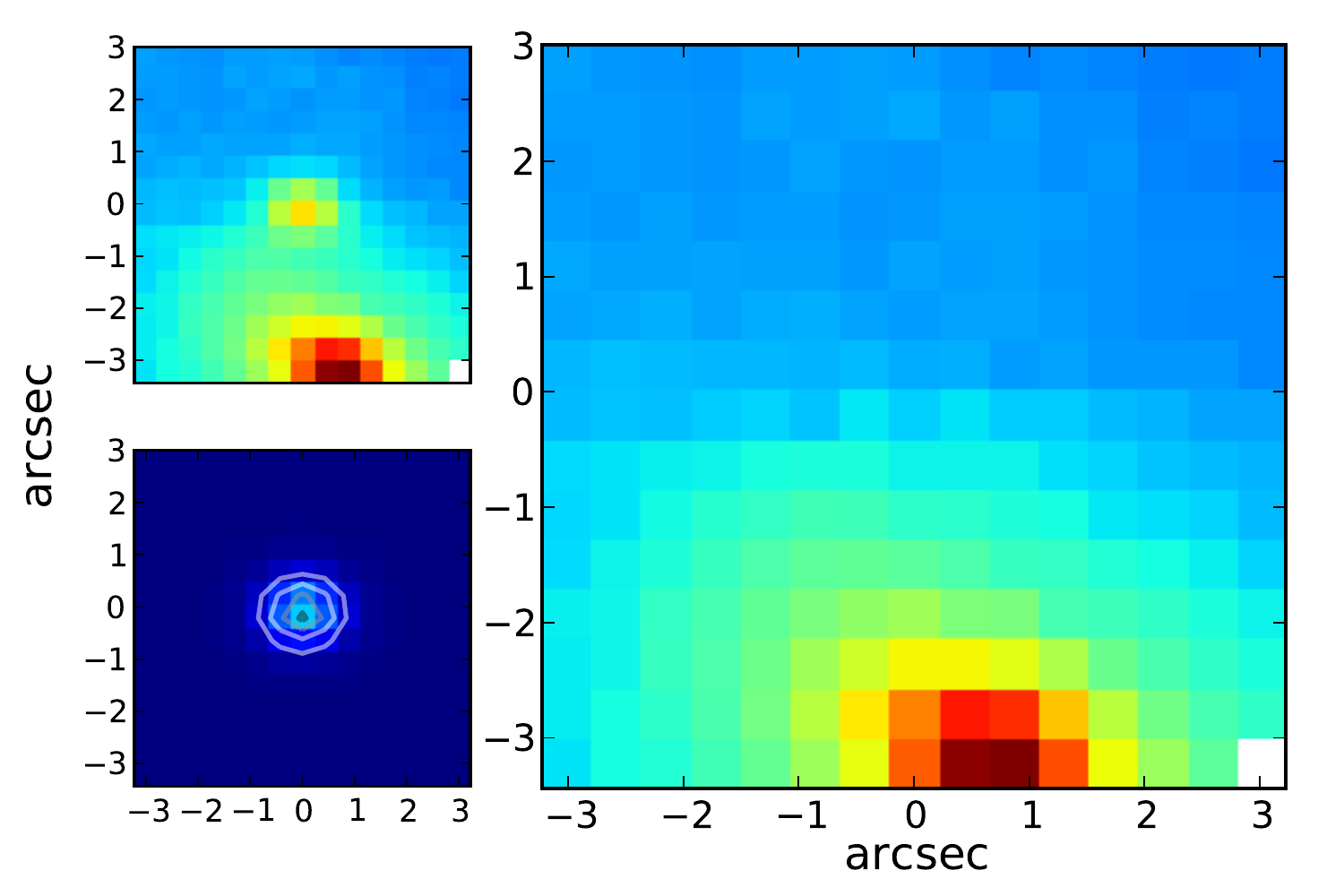}
  \caption{SN subtraction from a host cube.  Different reconstructed
    images, made by integrating $x,y,\lambda$-cubes along the
    $\lambda$ axis, are shown for a single observation of
    SNF20070326-012. \emph{Top left:} Original-flux calibrated cube
    containing host, sky, and SN signals. \emph{Bottom left:} Pure SN
    cube obtained after host subtraction (image) and fitted PSF
    model (contours).  \emph{Right:} SN-subtracted host cube.  }
  \label{fig:DDT-subtract-psf}
\end{figure}

For the present local host property analysis, we went one step
further and subtracted the aforementioned fitted PSF from the
original cubes to obtain \emph{SN-subtracted} host cubes.
Figure~\ref{fig:DDT-subtract-psf} shows an example of such a
subtraction procedure.  Since the SNfactory acquires SN~Ia time series,
this technique provides as many local host observations as SN
pointings (on average~13), in addition to the final references (on
average~2).

\subsubsection{Sky Subtraction}
\label{sec:sky-subtraction}

Since SNIFS has a FoV spanning only $\sim 6\arcsec$ across, it is usually
not possible to find a region entirely free of host signal, as is
usually done in photometry or long-slit spectroscopy.  The night sky
spectrum is a combination of the atmospheric molecular emission,
zodiacal light, scattered star light, along with the moon contribution
if any \citep{hanuschik_flux-calibrated_2003}.  We have developed a
sky spectrum model from a principal component analysis (PCA) on
700~pure sky spectra obtained from standard star exposures in various
observing conditions.  The $B$ and $R$ channels are analyzed
independently and in slightly different ways.

\begin{figure}
  \centering
  \includegraphics[width=\linewidth]{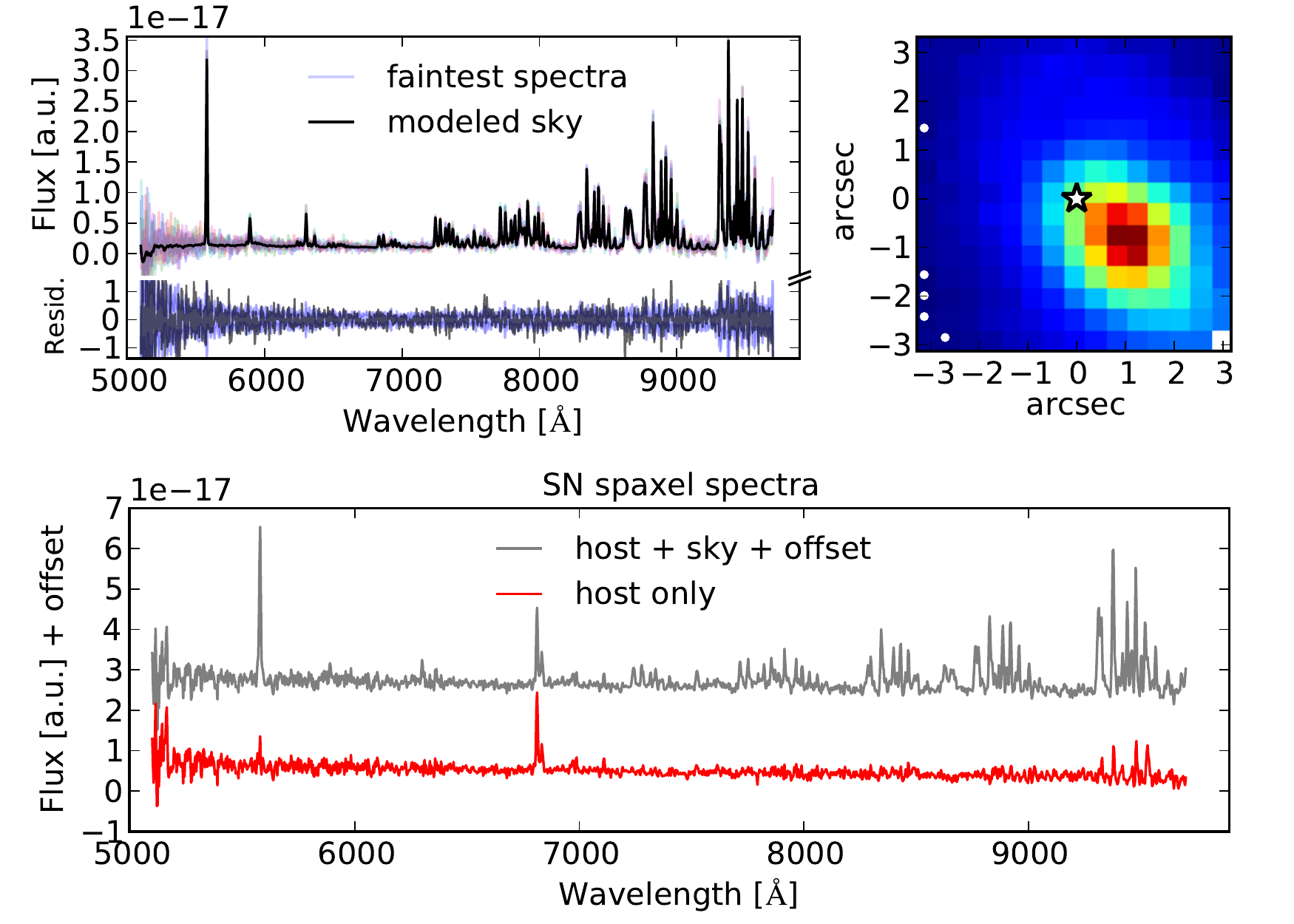}
  \caption{The sky subtraction process for an $R$ cube of
    host-SNF20060512-001.  \emph{Top right:} Reconstructed image of a
    final reference acquisition (it could as well be an SN-subtracted
    cube); the central star marker indicates the SN location.  The dot
    markers indicate the faintest spaxels used to measure the presumed
    sky.  \emph{Top left:} modeled sky (black) fit over the mean of
    the five faintest spectra (colored); the bottom part shows the fit
    residuals, compared to the presumed sky error (blue band).
    \emph{Bottom:} Host spectrum at the SN location before (gray line)
    and after (red line) the sky subtraction.  The emission at
    6830~\AA{} is the \Ha+[\ion{N}{ii}] gas line complex, left
    untouched by the sky subtraction procedure.  The few small
    features further to the red are residuals from the sky
    subtraction, but this part of the spectrum is not used in this
    analysis.  See Sect.~\ref{sec:Map_creation} for details.}
  \label{fig:Sky_example}
\end{figure}

The emission lines in $R$ sky spectra (mainly oxygen and OH bands) are
easily isolated from the underlying continuum.  This continuum is
fitted by a 4th-order Legendre polynomial over emission line-free
wavelength regions.  The PCA is then performed on the
continuum-subtracted emission line component.  Eight principal
components (PCs) are necessary to reconstruct the emission spectrum
with a median reduced $\chi^2$ of~1. An example of a red-sky
  spectrum fit using this 13-parameter model is given in
  Fig.~\ref{fig:Sky_example}.

The $B$ sky spectrum is dominated by diffuse light and shows
absorption features, as well as a few Herzberg O$_2$ lines in the
bluer part of the spectrum.  For this channel a PCA is performed
  directly on the observed sky spectra, since it was not possible to
  disentangle the different physical components as easily as for the
  $R$ sky spectra.  We find that four PCs are necessary to reach a
median reduced $\chi^2$ of~1.

In total, our $B+R$ sky model is therefore described by
$4+(5+8)=17$~linear parameters.  Since this model is trained on
selected pure sky spectra, none of the PCs should mimic galactic
features: even if some chemical elements are common (e.g. the hydrogen
or nitrogen lines), host galactic lines are redshifted, and cannot be
spuriously fit by the model.

Each spaxel of the SN-subtracted cube can be decomposed as follows:
\begin{equation}
  \label{eq:spaxel_composition}
  \mathrm{spaxel}(x,y,\lambda) =
  \mathrm{host}(x,y,\lambda) + \mathrm{sky}(\lambda),
\end{equation}
where the spatially flat sky component does not depend on the spaxel
location $(x,y)$.  We define the \emph{presumed} sky as the mean
spectrum of the 5~faintest spaxels, i.e. with the smallest host
contribution.  Since the $\mathrm{host}(x,y,\lambda)$ component of
Eq.~\ref{eq:spaxel_composition} may not be strictly null for these
spaxels, the \emph{presumed} sky may still contain galactic features,
e.g. hydrogen emission lines, and cannot be used directly as a valid
estimate of the true sky spectrum.  We therefore fit the presumed sky
spectrum with our model: the resulting \emph{modeled} sky is free from
any galactic features, and can safely be subtracted from each spaxel
to obtain a pure host cube. (See Fig.~\ref{fig:Sky_example} for an
example on $R$-channel.)

This procedure could slightly overestimate the sky continuum in cases
of bright host signal even in the faintest spaxels, and ultimately
lead to a under-estimation of the \Ha{} flux measurements
\citep{groves_balmer_2012}.  This effect is however insignificant in
comparison to our main source of \Ha{} measurement error, related to
our inability to correct for host dust extinction (see
Sect.~\ref{sec:Map_creation}).

\subsubsection{Spectral Merging}
\label{sec:merge-cubes}

In this analysis, we focus on the host properties of the SN local
environment, which we define has having projected distances less than
1~kpc from the SN.  This radius was chosen since it is greater than
our median seeing disk for our most distance host galaxies, at
$z=0.08$. ($1\arcsec = 1.03$~kpc at $z=0.05$.) No allowance is made
for host galaxy inclination in defining this local region.

Once the SN and the sky components have been subtracted, we compute
from each cube the mean host spectrum within 1~kpc around the SN.
This requires precise spatial registration (a by-product of the 3D
deconvolution algorithm, see Sect.~\ref{sec:SN-subtraction}) and
atmospheric differential refraction correction.  Those spectra (15 on
average per SN) are then optimally averaged and merged to get one host
spectrum of the local environment per SN.  The spectral sampling of
these merged-channel spectra is set to that of the $B$-channel,
i.e. 2.38~\AA.  In the same fashion, we are able to combine 3D-cubes
to create 2D-maps of any host property.

The spectra are corrected for Milky-Way extinction
\citep{schlegel_maps_1998}.  In this paper, observed fluxes are
expressed as surface brightnesses, and wavelengths are shifted to
rest-frame.  We only consider the 89 SNe~Ia in the main SNfactory
redshift range $0.03 < z < 0.08$ for spatial sampling reasons.
Namely, at lower $z$ the final SNIFS field of view remaining
  after spectral merging often subtends a radius of less than 1~kpc
  surrounding the SN location, while at higher $z$ the typical seeing
  disk subtends substantially more than 1~kpc. Since this selection
is based on redshift only, it does not introduce bias with respect to
host or SN~Ia properties.  \citep[See][ who showed that our host data
follow regular galaxy characteristics.]{childress_data_2013}

The influence of the PSF, i.e. the variation of the amount of
independent information with redshift for a given kpc-radius aperture,
is discussed in Sect.~\ref{sec:Assumption_discussion}, where we show
that our analysis is free from any redshift bias.

\subsection{Measurement of the Local Host Properties}
\label{sec:Map_creation}

\begin{figure}
  \centering
  \includegraphics[width=\linewidth]{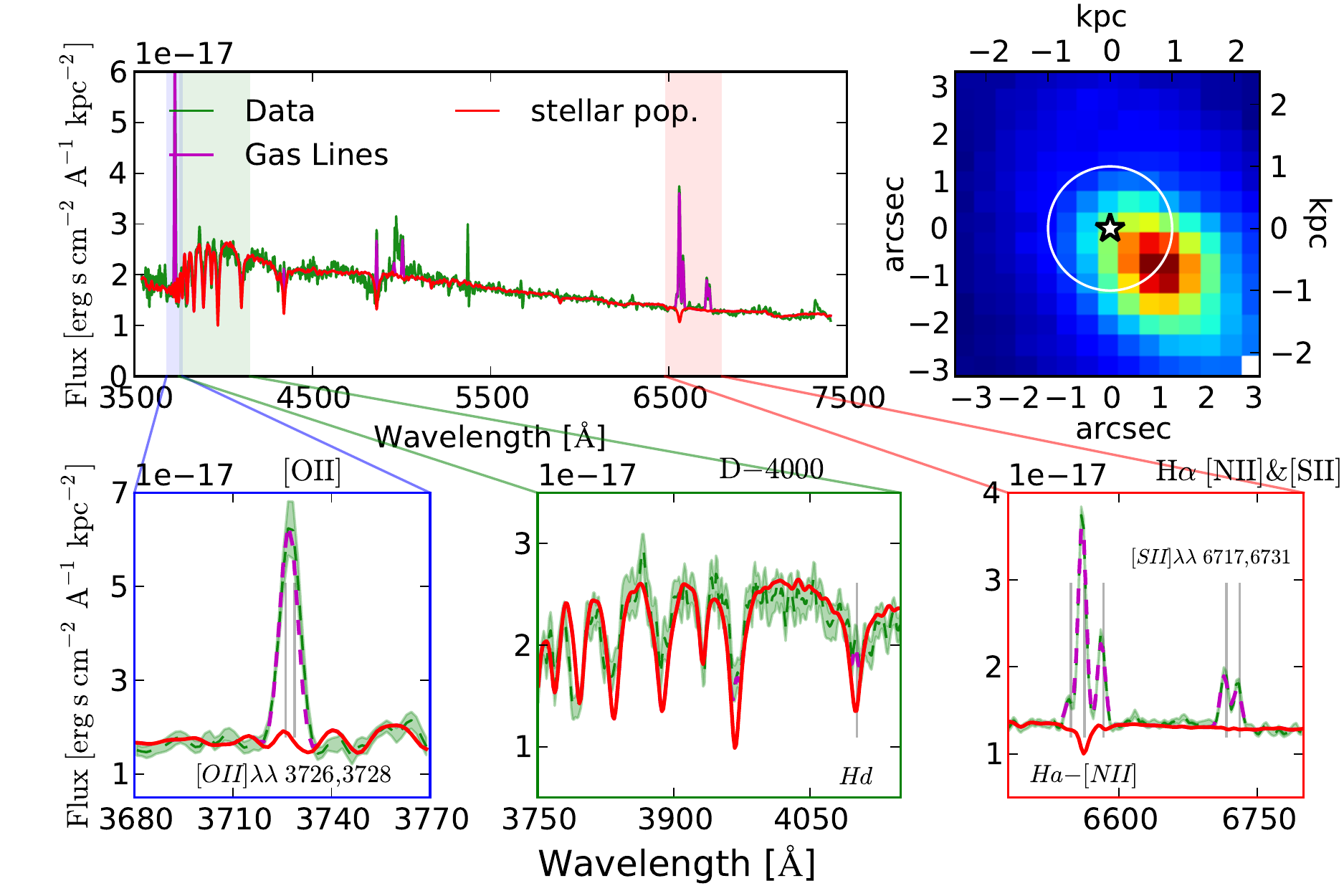}
  \caption{Spectral fit of the host mean spectrum in the projected
    1~kpc radius around SNF20060512-001.  \emph{Top right:}
    Reconstructed image of a final reference exposure.  The circle
    aperture shows the 1~kpc aperture centered at the SN location
    (central star marker).  \emph{Top left:} mean host spectrum within
    1~kpc radius around the SN (rest-frame, green line), averaged over
    21 cubes (including two final references).  The best ULySS fit is
    shown as the sum of a stellar continuum (red line) and gas lines
    (magenta line).  \emph{Bottom, from left to right:} Zoom on
    [\ion{O}{ii}]$\lambda\lambda3726,3728$, 4000 \AA{} break, and
    \Ha{} + [\ion{N}{ii}]$\lambda\lambda6548,6584$ +
    [\ion{S}{ii}]$\lambda\lambda6716,6731$ wavelength regions.  }
  \label{fig:ULySS_zoom_fit}
\end{figure}

A galactic spectrum is a combination of a continuum component with
absorption lines mostly from the stars, and emission lines from the
interstellar gas.  In this paper, we restrict our analysis to the
\Ha{} signal around the SN.  Since hydrogen emission line intensities
can be affected by underlying stellar absorption, it is necessary to
estimate the stellar continuum to obtain accurate \Ha{}
measurements. (See Fig.~\ref{fig:ULySS_zoom_fit}.)

The \emph{University of Lyon Spectroscopic analysis
  Software}\footnote{http://ulyss.univ-lyon1.fr}
\citep[ULySS,][]{koleva_spectroscopic_2008, koleva_ulyss:_2009} is
used to disentangle the stellar and gas components.  It fits
simultaneously the following three components: stellar populations
from the MILES library
\citep{sanchez-blazquez_medium-resolution_2006}; a set of emissions
lines -- Balmer's hydrogen series in addition to
[\ion{O}{ii}]$\lambda\lambda 3726,3728$, [\ion{O}{iii}], [\ion{N}{ii}]
and [\ion{S}{ii}]; and a multiplicative \emph{ad~hoc} continuum
correcting for both internal dust extinction and any remaining
large-scale flux mismatch between the data and the template library.
Only wavelengths matching the MILES library rest-frame are considered
(3540--7410~\AA), and gas emission lines all share a common redshift
and velocity dispersion.  Because of this added constraint, gas
emission flux uncertainties given by ULySS cannot be trusted readily.
We therefore ultimately fit gaussian profiles on stellar-corrected
emission lines to derive \Ha{} fluxes and accurate uncertainties.

Figure~\ref{fig:ULySS_zoom_fit} presents the fit of the mean spectrum
of the local environment for SNF20060512-001.  In this particular case
21~exposures per channel, including two final references, have been
combined.  The stellar component is strongly constrained by the D4000
feature, and thereby gas flux measurements properly account for
stellar absorption.

In this paper, we focus our study on the \Ha{} line as a tracer of the
star-forming activity \citep{kennicutt_star_1998}.  The spectra are
not corrected for the internal host extinction, because the Balmer
decrement method \citep{calzetti_dust_1994} cannot be reliably
applied: the H$\gamma$ lines are usually too faint, and H$\beta$ lines
happen to lie at the SNIFS dichroic cross-over ($4950 < \lambda <
5150$~\AA{} in the observer frame, see Fig.~\ref{fig:ULySS_zoom_fit}),
where the S/N is low and fluxes harder to calibrate.

\section{Local H$\alpha$ surface brightness analysis}
\label{sec:Analysis}

The \Ha{} emission within a 1~kpc projected radius can be conveniently
recast as a surface brightness, which we will designate with \SigmaHa.
Because the measurement is projected, it actually represents the total
\Ha{} emission in a 1~kpc radius column passing through the host
galaxy centered on the SN position.  Thus, it is possible that \Ha{}
we detect, in whole or in part, is in the foreground or background of
the 1~kpc radius sphere surrounding the SN.

However, there are several factors that greatly suppress projection
effects for \Ha{} relative to stars.  One is that \Ha{} depends on the
square of the electron density, so most \Ha{} comes from those rare
regions with high gas density.  This advantage is compounded by the
fact that young stars capable of ionizing hydrogen also occur
preferentially in clumps and regions of higher gas density.
\citep[see][especially section 2.2, for a detailed discussion of
\ion{H}{ii}~regions and their clumpiness]{kennicutt_star_2012}.  In
addition, while stars are spread across thin and thick disks, bulges
and halos, \ion{H}{ii}~regions are concentrated in the thin galactic
disk, where a typical scale height is only $\sim0.1$~kpc
\citep{paladini_spatial_2004}.  Thus the line-of-sight depth extends
beyond our canonical 1~kpc radius only for very high inclinations
(i.e., $i > 84\deg$).  Finally, the \Ha{} covering fraction is
generally small because \ion{H}{ii}~regions are so short-lived
\citep[see][for a detailed review of tracers of star formation
activity.]{calzetti_star_2012}

In the following analysis, a few cases (7/89) of SNe with locations
projected close to the core of very inclined hosts have been set aside
because of the high probability of such misassociation (designated as
``p.m.s.'', see Table~\ref{tab:subgroups}).  In
Sect.~\ref{sec:Assumption_discussion} we show that these cases have no
influence on our results.

A different projection effect occurs if SNe~Ia travel far from their
formation environment before they explode.  Due to random stellar
motions this will occur, and its importance will itself depend on
SNe~Ia lifetimes and the host velocity dispersion.  If as expected,
tardy SNe~Ia are associated with old stellar populations, this
correspondence may be lost only in cases where the host is globally
star forming -- i.e. for spiral rather then elliptical galaxies --
{\it and} the SN motion has projected it onto a region of active star
formation.  As with the line-of-sight projection discussed above, this
situation should be uncommon.  On the other hand, prompt progenitors
should not have drifted from their stellar cohort by more than our
fiducial 1~kpc radius.  This cohort is likely to have been an open
cluster or OB association; open clusters remain bound on the relevant
timescale, while OB associations have very low velocities dispersions
\citep[$\sim 3~\mathrm{km\,s^{-1}}$][]{dezeeuw_1999, pz_2010,
  roser_2010}.  Consequently, there should be little contamination due
to this type of projection effect for young progenitors.

Thus, the net effect will be that for some old SN progenitors in
globally star-forming galaxies a false positive association with \Ha{}
may arise due to these projection effects.  However, we expect the
number of such cases to be small, as discussed later in this section.

For ease in interpretation, we split our sample in two, using the
median \SigmaHa{} value of our distribution, $\log(\SigmaHa{})=
38.35$, as the division point.  This corresponds to a star formation
surface density of $1.22\times
10^{-3}\;\mathrm{\Msol\,yr^{-1}\,kpc^{-2}}$ assuming Eq.~5 of
\cite{calzetti_calibration_2010}.  Cases where no \Ha{} signal was
found are arbitrarily set to $\log(\SigmaHa{})= 37$ in the figures.

The median \SigmaHa{} value happens to correspond to the threshold at
which we achieve a minimal $3\sigma$-level measurement sensitivity.
It also corresponds to the mean $\log(\SigmaHa{})$ of the warm
interstellar medium (WIM) distribution reported by
\cite{oey_survey_2007}.  The source of the WIM is controversial.  The
long-standing interpretation has been that the WIM is due to Lyman
continuum photons that have escaped from star-forming regions and then
ionize the ISM.  In this interpretation the WIM would still trace star
formation, but its distribution would be more diffuse than an
\ion{H}{ii}~region.  This distinction would not matter given our large
metric aperture since even in the ISM the optical depth to Lyman
continuum photons is large.  Recently, however, \citet{seon_witt_2012}
have interpreted the WIM as due to reflection of light from
\ion{H}{ii}~regions off of ISM dust.  In this case WIM \Ha{} would
still arise from star formation, but possibly from distances well
outside our metric aperture since the optical depth to \Ha{} photons
can be low.  By setting our division point above the nominal
$\log(\SigmaHa{})$ of the WIM we minimize the importance of these
details for our analysis.

We designate these two \SigmaHa{} subgroups as follows (see
Table~\ref{tab:subgroups}):
\begin{itemize}
\item 41~SNe~\sS{} with $\log(\SigmaHa) \geq 38.35$, which we refer to
  as ``locally star forming'',
\item 41~SNe~\sN{} with $\log(\SigmaHa) < 38.35$, which we refer to as
  ``locally passive''.
\end{itemize}

\begin{table*}
  \caption{Definition of the two \SigmaHa{} subgroups, and their
    respective weighted mean light-curve parameters.
    The ``p.m.s.'' column indicates the number of SNe with a
    potential misassociation of a \Ha{} detection due to projection effects,
    which have been excluded from the main analysis.
  }
  \label{tab:subgroups}
  \centering
  \begin{tabular}{llcccccc}
    \hline\hline
    Label & Group & $\log(\SigmaHa)$ & SNe &
    $\langle x_1 \rangle$ & $\langle c \rangle$ &
    $\langle\cHR \rangle$ [mag] & p.m.s.\\[0.4ex]
    \hline\\[-2ex]
    \sN & \emph{Locally} passive         & $< 38.35$ & 41 &
    $-0.29\pm 0.14$ & $0.029\pm 0.010$ &
    $-0.039\pm0.023$ & 2 \\[0.1ex]
    \sS  & \emph{Locally} star-forming & $\geq   38.35$ & 41 &
    $+0.08 \pm 0.10$ & $0.065\pm 0.013$ &
    $+0.056\pm0.020$  &  5 \\
    \hline
  \end{tabular}
\end{table*}

\begin{figure}
 \centering
 \includegraphics[width=\linewidth]{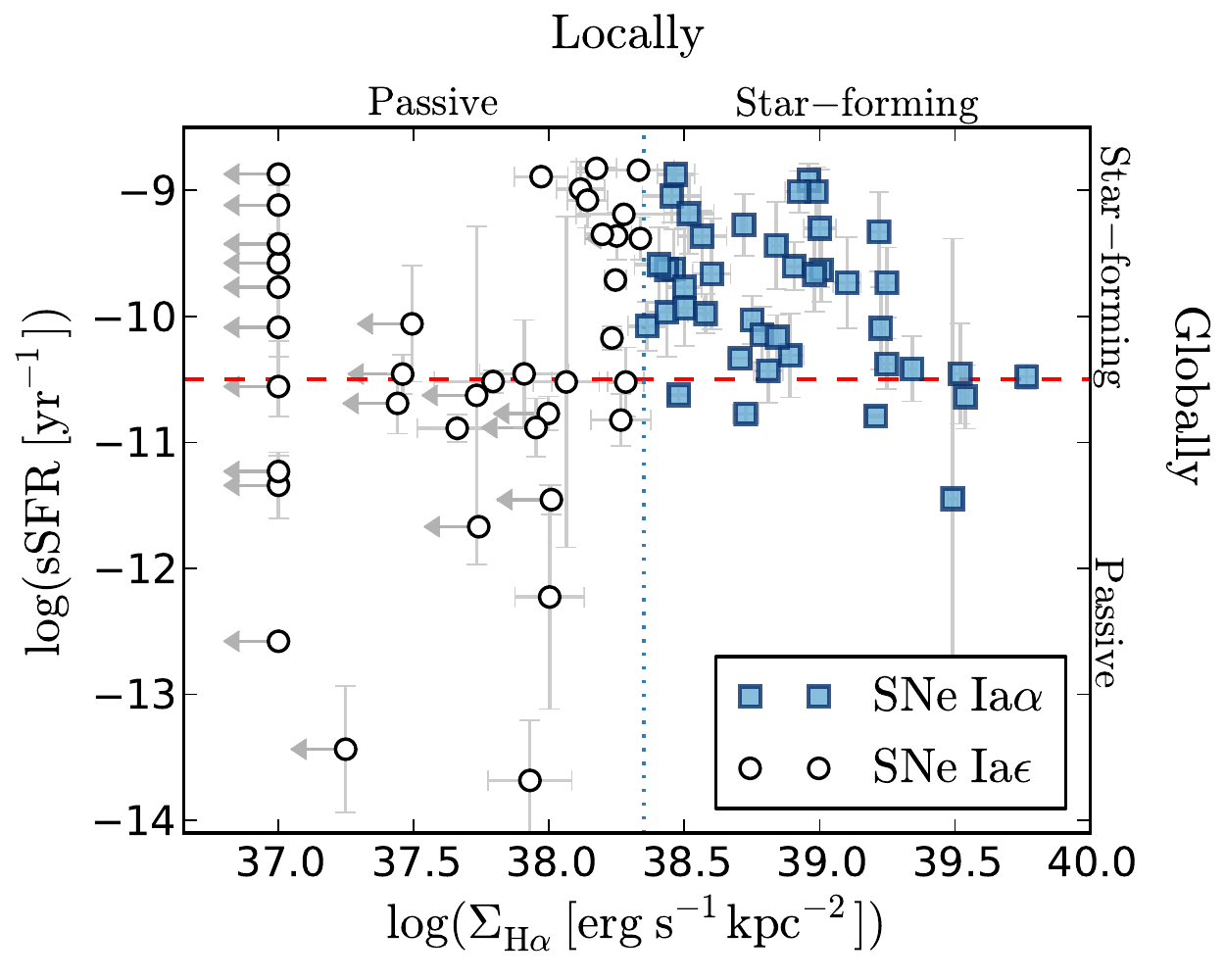}
 \caption{Local vs. global star-forming environment.  Blue squares and
   open-circles show the SNe~Ia from locally star-forming and passive
   regions, respectively ($\log(\SigmaHa) \gtrless 38.35$, blue-dotted
   line).  Gray arrows indicate points with \SigmaHa{} signals
   compatible with zero (detection at less than $2\sigma$).  The
   red-dashed line indicates the limit that is commonly used to
   distinguish between star-forming and passive galaxies when using
   global host galaxy measurements ($\log(\mathrm{sSFR}) \gtrless
   -10.5$).  Three SNe~Ia from our sample do not have sSFR
   measurements. }
  \label{fig:LHa_ssfr}
\end{figure}

Figure~\ref{fig:LHa_ssfr} presents a quantitative picture of the point
made visually in Fig.~\ref{fig:global_vs_local} by illustrating the
non-trivial relationship between local and global measurements for
cases considered passive.  The global quantities for our hosts are
taken from our compilation in \citet{childress_data_2013}; in three
cases global sSFR measurements are unavailable for SNe~Ia in our local
environment sample.  Figure~\ref{fig:LHa_ssfr} shows that half of the
SNe with locally passive environments (20/38) have a globally
star-forming host (traditionally defined as $\log(\mathrm{sSFR}) >
-10.5$).  This highlights the existing degeneracy when employing
global properties: the SNe~Ia hosted in globally star-forming galaxies
can have \emph{either} star-forming or passive local environments
whereas SNe hosted in globally passive galaxies are almost certain to
have a locally passive environment.

SNe~\sS, having \SigmaHa{} above the median value, are most likely to
have young progenitors since an \Ha{} signal has been positively
detected in their vicinities.  The large number of SNe~Ia with
detections of local star formation in Fig.~\ref{fig:LHa_ssfr} itself
provides an indication that young progenitors exist, in agreement with
the statistical analysis of \cite{aubourg_evidence_2008}.  On the other
hand, the wide range of observed \SigmaHa{} is an indication of SNe
from both passive and star-forming regions, i.e. from both young and
old progenitors \citep{scannapieco_type_2005, mannucci_supernova_2005,
  mannucci_two_2006, sullivan_rates_2006}.

In Fig.~\ref{fig:LHa_ssfr} there is a relative dearth of SNe~Ia with
\Ha{} detections for $\log(\SigmaHa) < 37.9$ in hosts that are
globally star-forming.  This then highlights a population having $38.0
< \log(\SigmaHa) < 38.35$ in globally star-forming hosts which are
counted as locally passive when we split our sample.  Their proximity
to the typical WIM level suggests that these could be cases of SNe~Ia
from old progenitors whose \SigmaHa{} value is boosted by projection
onto the WIM in their hosts.  But, they equally well may be SNe~Ia
from young/intermediate age progenitors where strong star formation
has already ebbed at that location in their host.  In
Sect.~\ref{sec:Assumption_discussion} we show that moving these cases
into the star-forming sample does not change our results
significantly.  Therefore we prefer to employ a simple split since it
is not tuned to any patterns in the data.

\section{\SigmaHa{} and SN Ia Light-Curve Properties}
\label{sec:Results}

We will now compare $\log(\SigmaHa{})$ to the SN light-curve
parameters (Sect.~\ref{sec:Groups_variations}) and to the corrected
Hubble residuals \cHR{} (Sect.~\ref{sec:HR_results}).  We show that SNe
from locally passive regions have faster light curve decline rates,
that the color of SNe~Ia is correlated with \SigmaHa{}, and that there
is a significant magnitude offset between SNe from locally
star-forming (\sS) and locally passive (\sN) regions.

The SALT2 lightcurve parameters $c$ and $x_{1}$ are known to correlate
with the uncorrected absolute $B$-magnitude at maximum light, $M_B$.
The dispersion in $M_B$ can be reduced significantly by taking
advantage of these correlations \citep{guy_salt2:_2007}:
\begin{equation}
  \label{eq:cHR}
  \cHR \equiv (M_B-M_B^{0}) + (\alpha{} x_{1} - \beta c),
\end{equation}
where the stretch coefficient $\alpha$, the color coefficient $\beta$,
and the average absolute magnitude of the SNe~Ia, $M_B^{0}$, are
simultaneously fit over the SN sample during the standardization
process.  The SALT light curve parameters and Hubble residuals for our
SNe~Ia have been presented previously in \cite{bailey_using_2009},
\cite{chotard_reddening_2011}, and \cite{childress_host_2012}.

For the statistical analyses that follow we will use the
Kolmogorov-Smirnov test (KS-test) to estimate the discrepancy between
observed parameter distributions ($x_{1}$, $c$, and \cHR{}).  We also
will compare the weighted-mean values of those distributions between
the \SigmaHa{} groups, which we denote using the form $\delta\langle
X\rangle(A-B)\equiv \langle X\rangle_A - \langle X\rangle_B$.  When
exploring correlations we will employ the non-parametric Spearman rank
correlation test, quoting the correlation strength, $r_S$, and
probability, $p_S$.  $p$-values below 5\% (e.g., $p_{KS}< 0.05$)
highlight statistically important differences between distributions,
and we will consider $p$-values below 1\% to be highly significant.

\subsection{Light-Curve Stretch and Color}
\label{sec:Groups_variations}

\begin{figure*}
  \centering
  \includegraphics[width=0.6\textwidth]{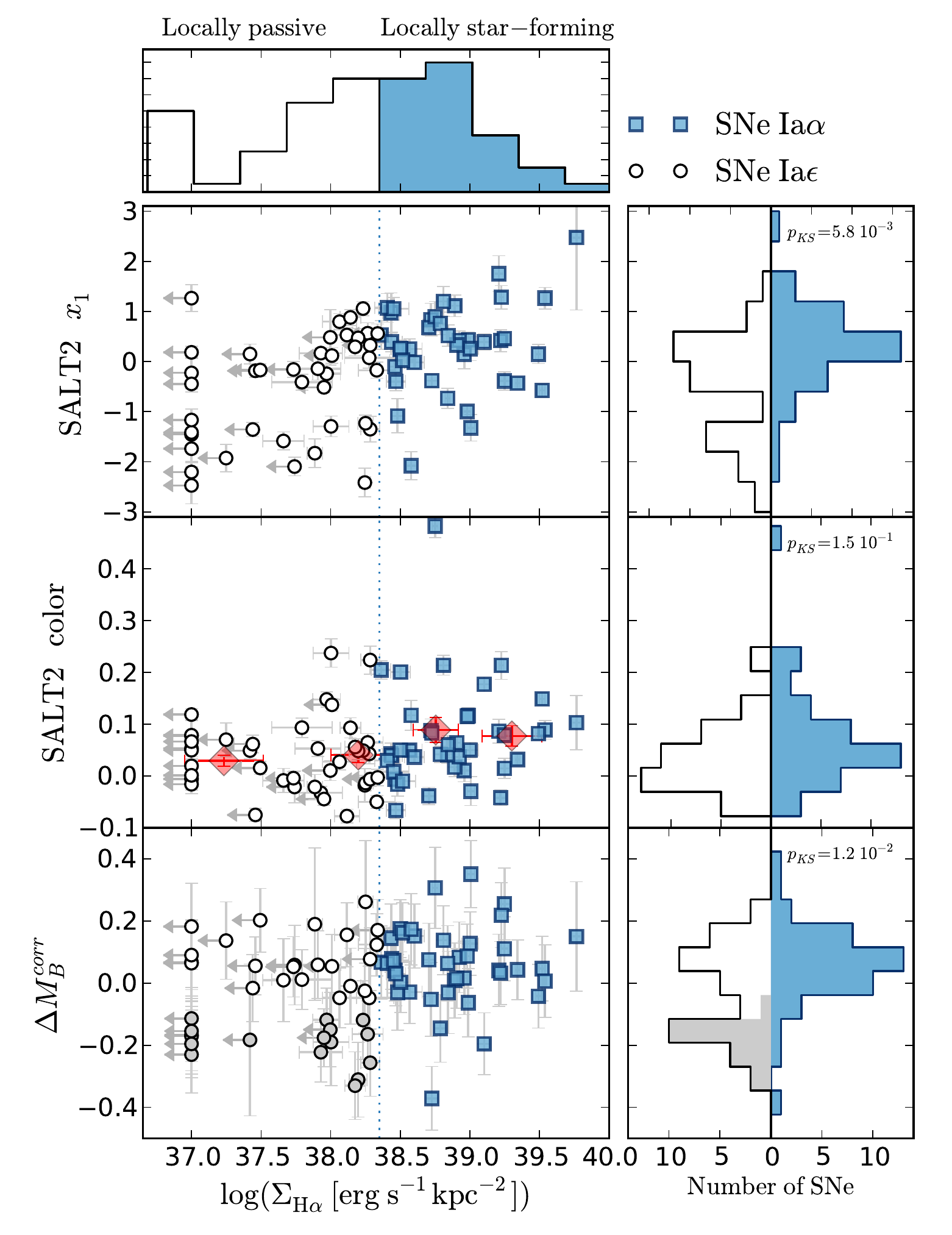}
  \caption{SN~Ia parameters as a function of $\log(\SigmaHa)$, the
    surface brightness in the 1~kpc radius around the SNe~Ia.
    \emph{Upper panel}: The $\log(\SigmaHa)$ distribution colored with
    respect to \SigmaHa{} subgroups: \sN{} and \sS{} are shown as
    empty-black and filled-blue histograms, respectively.  \emph{Main
      panels, from top to bottom:} SALT2 stretch $x_1$, color $c$, and
    the Hubble residuals corrected for stretch and color \cHR{} (with
    a standardization done over the whole sample).  Symbol shapes and
    colors are the same as in Fig.~\ref{fig:LHa_ssfr}.  In the middle
    panel, filled-red diamonds show the mean SALT2 color measured in
    bins of $\log(\SigmaHa)$: $x$-axis error bars indicate the
    $\log(\SigmaHa)$ dispersion of the SNe of the considered bin.  In
    the \cHR{} panel, the \Mtwo{} SNe (\sN{} with \cHR{}<-0.1, see
    text) are shown as filled-gray circles.  \emph{Right panels, from
      top to bottom}: Marginal distributions of $x_1$, $c$, and \cHR{}
    by subgroups.  These bi-histograms follow the same color code as
    the $\log(\SigmaHa)$ histogram.  The contribution of the \Mtwo{}
    SNe to the \sN{} \cHR{} distribution is shown in gray in the lower right panel. }
  \label{fig:HaFrac__vs__salt2_HR}
\end{figure*}

The stretch parameter $x_{1}$ and color $c$ are shown in
Fig.~\ref{fig:HaFrac__vs__salt2_HR} as a function of $\log(\SigmaHa)$.
The weighted means and dispersions for the two subgroup are summarized
in Table~\ref{tab:subgroups}.

\subsubsection{The Light-Curve Stretch}
\label{sec:stretch_FracHa}

The SALT2 stretch $x_1$ and the $\log(\SigmaHa{})$ values are
correlated, as evidenced by a highly significant Spearman rank
correlation coefficient: $r_s=0.42$, $p_s=7.4\times 10^{-5}$,
$4.2\sigma$.  This shows that SNe~Ia from low \SigmaHa{} environments
have on average faster light-curve decline rates.  When comparing
\SigmaHa{} subgroups, we find $\delta\langle x_1\rangle(\sS-\sN)=0.37
\pm 0.17$ ($p_{KS} = 5.8\times 10^{-3}$).

\begin{figure*}
  \centering
  \includegraphics[width=\textwidth]{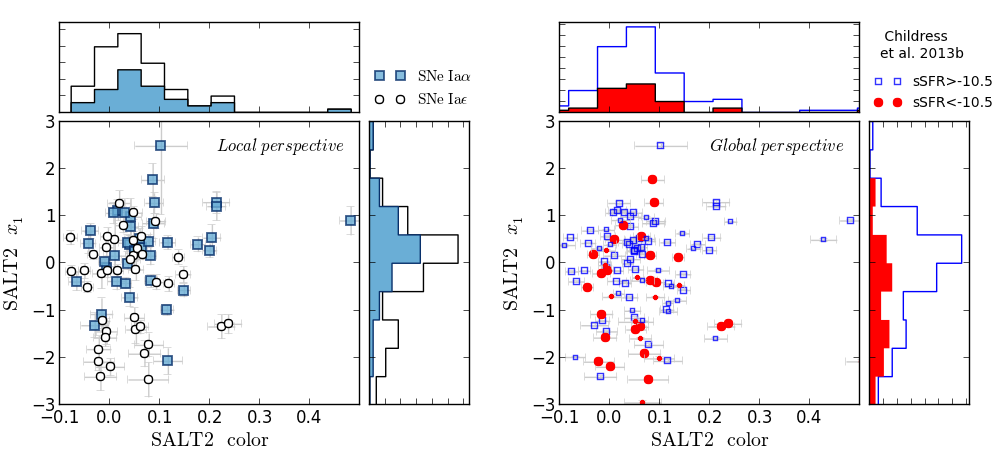}
  \caption{SALT2 stretch $x_1$ vs. color $c$.  \emph{Left:} The local
    perspective.  SNe markers and $x_1$ and $c$ marginalized stacked
    distributions follow the same color-shape code as
    Fig.~\ref{fig:HaFrac__vs__salt2_HR}.  \emph{Right:} The global
    perspective.  Open-blue squares show the globally star-forming
    hosts and red dots show the globally passive hosts
    ($log(\mathrm{sSFR})\lessgtr -10.5$ respectively).  Big markers
    indicate SNe with local measurements, while small markers denote
    SNe not in our local sample (failed redshift cut, but passed the
    quality cuts, see Sect.~\ref{sec:SNIFS}).  Stacked histograms are
    the marginalized distributions of $x_1$ and $c$ using the marker
    color codes.  These figures compare to Fig.~2 of
    \cite{lampeitl_effect_2010}.  Notice the mix of local environments
    for moderate $x_1$ while low $x_1$ are almost exclusively from
    passive environments.}
  \label{fig:HaFrac__X1_vs_C}
\end{figure*}

At first this might appear to be just the same as the well-established
effect whereby globally passive galaxies are found to host SNe~Ia with
lower stretch while star-forming host galaxies dominate at moderate
stretches \citep{hamuy_search_2000, neill_local_2009,
  lampeitl_effect_2010}.  Indeed, the $x_1$ distributions of SNe in
globally passive and in globally star-forming hosts
($\log(\mathrm{sSFR}) \lessgtr -10.5$ \footnote{See
  \cite{childress_data_2013} for details on the measurement of sSFR
  for the SNfactory sample.}) differ significantly, having $p_{KS} =
2.0\times 10^{-3}$ for the SNfactory and $p_{KS} = 10^{-7}$ for SDSS
\citep{lampeitl_effect_2010}.  We compare in
Fig.~\ref{fig:HaFrac__X1_vs_C} the global and local pictures in the
$(c,x1)$ plane.  These two figures can be compared to Fig.~2 of
\cite{lampeitl_effect_2010} for the SDSS sample.

However, our analysis of the local environment displays quite a
different stretch distribution for passive environments.  We find that
40\% of SNe that are locally passive have moderate stretches.  This is
in strong contrast to global analyses.  For our SN sample only 25\% of
globally passive hosts have SNe~Ia with $x_{1} > -1$, and in the
global analysis of the SDSS dataset \cite{lampeitl_effect_2010} the
percentage of passive environments with moderate stretch is even
lower, a mere 11\%.  This suggests that many SNe~Ia originating in
passive environments have been mistakenly categorized as coming from
star-forming environments by analyses employing global host galaxy
properties.  Were we to lower our dividing line to $\log(\SigmaHa{}) =
38$, our global and local analyses would then agree, but this still
would not result in the paucity of passive environments at moderate
stretches seen in the SDSS sample.

Furthermore, we find even stronger evidence that SNe with fast decline
rates ($x_1<-1$) arise in passive environments.  In our local analysis
15/18, or 83\% are \sN{}, and all have a $\log(\SigmaHa) \leq 39$.  In
contrast, as Fig.~\ref{fig:HaFrac__X1_vs_C} shows, our global analysis
for $x_1<-1$ shows only mild dominance by globally passive
environments, with 60\% being globally passive.  In the SDSS sample of
\citet{lampeitl_effect_2010}, among the SNe~Ia with fast decline rates
only 54\% are globally passive.

In summary, we find that lower stretch SNe~Ia are hosted
almost exclusively in passive local environments, but that SNe~Ia
having moderate stretches are hosted by all types of environments.

\subsubsection{The Light-Curve Color}
\label{sec:colot_FracHa}

We also find that SNe~Ia from star-forming environments are on average
redder than those from passive environments at the $2.2\sigma$ level:
$\delta \langle c \rangle (\sS-\sN) = 0.036 \pm 0.017$ ($0.031 \pm
0.015$ when removing the reddest, SNF20071015-000).  This effect is
more significant when comparing the tails of the $\log(\SigmaHa)$
distribution (lower and upper quartiles), where we find distribution
differences with $p_{KS}= 0.026$.

As seen in Fig.~\ref{fig:HaFrac__vs__salt2_HR}, the SALT2 color, $c$,
and the $\log(\SigmaHa{})$ values are correlated, with a $2.0\sigma$
non-zero Spearman rank correlation coefficient $r_s=0.21$, with $p_s=
0.054$.

Since dust is associated with active star formation
\citep[e.g.][]{charlot_simple_2000}, a color excess is expected for
SNe~Ia in strong \SigmaHa{}~environments. \citep[See][for a discussion
of the Type~Ia color issue.]{chotard_reddening_2011} This trend has
been suggested for SNLS SNe in \cite{sullivan_dependence_2010} but is
not seen in SDSS data \citep{lampeitl_effect_2010}.

As discussed earlier, by analyzing integrated galaxy properties,
global host studies merge SNe~\sN{} with $x_1>-1$, and the SNe~\sS{},
into a single star-forming host category.
Figure~\ref{fig:HaFrac__X1_vs_C} shows that low and moderate stretch
\sN{} share the same color distribution.  It is then difficult for
global analyses to detect the color variation between the globally
star-forming and passive hosts, since both kinds harbor some \sN{}.
Indeed, in the SNfactory sample globally passive and globally
star-forming hosts produce SNe~Ia having the same SALT2 color
distribution ($p_{KS}=0.654$).

\subsection{Hubble Residuals}
\label{sec:HR_results}

The SALT2 raw Hubble residuals $\Delta M_{B}$ are corrected for their
stretch and color dependencies (\cHR{}) and are shown in
Fig.~\ref{fig:HaFrac__vs__salt2_HR} as a function of $\log(\SigmaHa)$.
The weighted mean values per \SigmaHa{} group are summarized in
Table~\ref{tab:subgroups}.

\subsubsection{The H$\alpha$ bias}
\label{sec:Halpha_bias}

We find with a $3.1\sigma$ significance that SNe from locally passive
environments are brighter than those SNe~Ia in locally star-forming
environments after color and stretch correction: $\DcHR = -0.094 \pm
0.031$~mag ($p_{KS}=0.012$).  The amplitude of this significant
brightness offset -- hereafter called the ``\Ha{}~bias'' -- is a
concern for precision cosmology, since the fraction of SNe~Ia from
passive environments most likely evolves with redshift, thus creating
a redshift-dependence on the average SNe~Ia luminosity.  Distance
measurements to these standardizable candles are therefore biased.  A
more complete discussion of the potential impact on cosmological
measurements is presented in Sect.~\ref{sec:Cosmo_consequences}.

\subsubsection{The Corrected Hubble Residual Distributions}
\label{sec:bimodality}

This leads us to investigate the possible origin of the brightness
offset between the SNe~\sS{} and the SNe~\sN{} (\Ha{}~bias) by
examining the \cHR{} distributions of each subgroup.  In
Fig.~\ref{fig:HaFrac__vs__salt2_HR} we notice the presence of a
subgroup of significantly brighter SNe hosted exclusively in locally
passive environments.  We thus distinguish two modes in the
distribution of corrected Hubble residuals: (1) The first mode exists
in all environments and has $\cHR\ga 0$.  (2) The second mode is
exclusive to SNe~Ia from low \Ha{} environments (\sN) and populates
the brighter part of the diagram almost exclusively: among the 20
SNe~Ia with $\cHR{} < -0.1$, 17~are \sN{}, and only one has a
$\log(\SigmaHa) > 39$.  For analysis purposes we will define and label
these modes as follows:
\begin{itemize}
\item \emph{Mode~1} (\Mone): 65~SNe consisting of all SNe~\sS{} along
  with SNe~\sN{} that have $\cHR>-0.1$~mag.  \vskip 5pt
\item \emph{Mode~2} (\Mtwo): 17~SNe~\sN{} with $\cHR < -0.1$~mag.
\end{itemize}
Except for the fact that both are from locally passive environments,
the \Mtwo{} subgroup and the subgroup of 15 fast declining
($x_1<-1$) SNe~Ia turn out to not be directly related, having
only 6~SNe in common.

\begin{figure}
  \centering
  \includegraphics[width=0.8\linewidth]{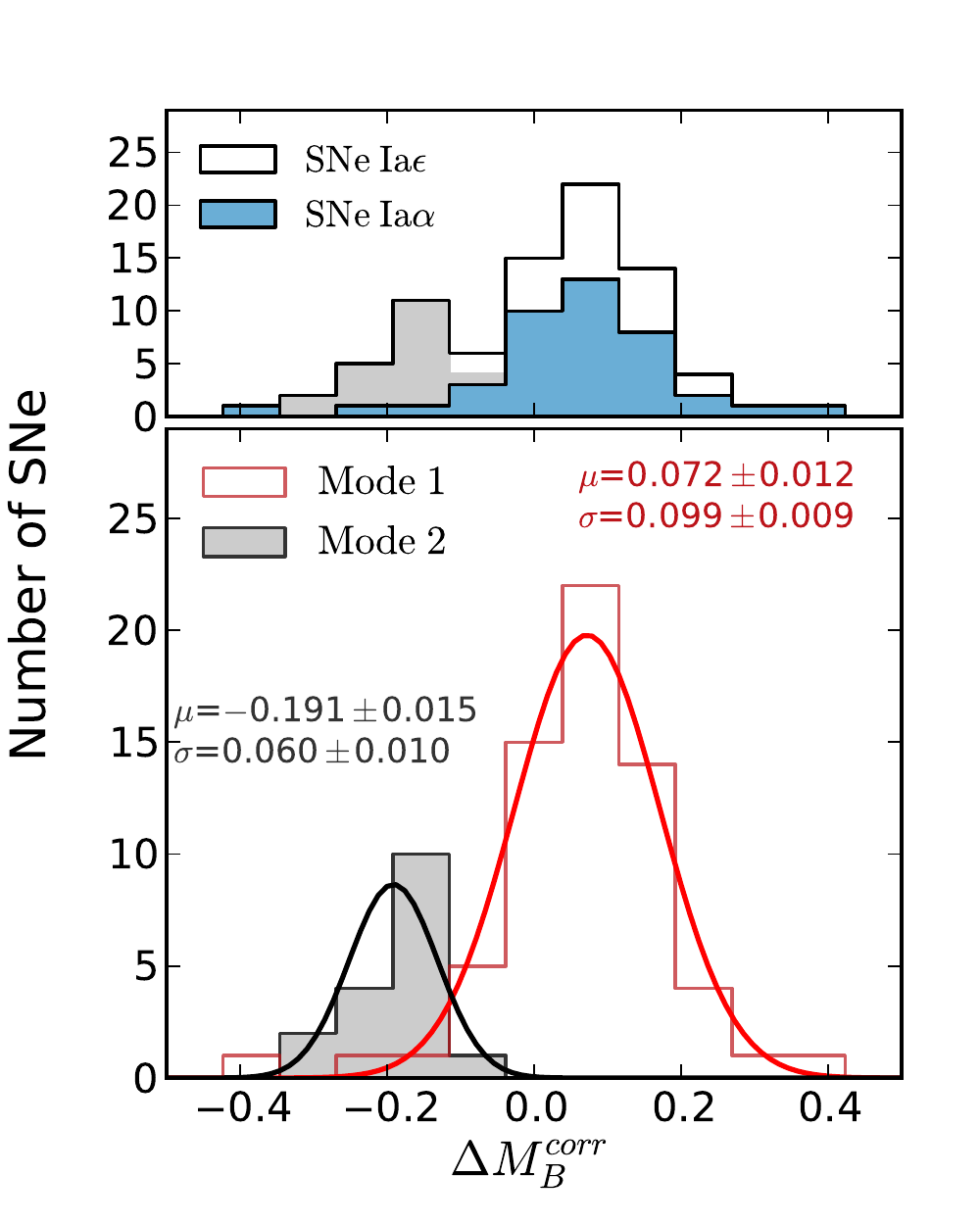}
  \caption{Distribution of corrected Hubble residuals per \SigmaHa{}
    group (\emph{top}, as in Fig.~\ref{fig:HaFrac__vs__salt2_HR}) and
    per mode (\emph{bottom}) for the full sample.  In the lower
    panel, the open-red and the filled-gray histograms represent the
    corrected Hubble residual distributions of \Mone{} and \Mtwo{},
    respectively.  By definition of the modes, the red histogram
    corresponds to the stack of the blue distribution with the open
    histogram above $\cHR=-0.1$ (see the top-panel), while the
    filled-gray histogram corresponds to the part of the open
    histogram below $\cHR=-0.1$.  Colored solid lines (red for \Mone{}
    and black for \Mtwo{}) show the best fit by a normal distribution
    (with $3\sigma$ clipping).  Fit parameters are quoted beside each
    gaussian. }
  \label{fig:cHR__Two_modes}
\end{figure}

Figure~\ref{fig:cHR__Two_modes} shows the distribution of these two
modes fitted by normal distributions (with $3\sigma$ clipping).
\Mone{} SNe are on average fainter ($\langle\cHR\rangle_{\Mone} =
0.072\pm0.012$~mag) and \Mtwo{} SNe~Ia are brighter
($\langle\cHR\rangle_{\Mtwo}= -0.191\pm0.015$~mag) than the full
sample.  The standard deviations are $0.099 \pm0.009$~mag and
$0.060\pm0.010$~mag for \Mone{} and \Mtwo, respectively.  The weighted
RMS gives similar results, with variations at a tenth of the error
level.  With no clipping, one additional, bright SN is then retained
in \Mone{}, raising the \Mone{} standard deviation to
$0.113\pm0.010$~mag.

These differences in $\langle\cHR\rangle$ would appear to be
extremely significant.  However in this case we chose the
dividing point in \cHR{} and thus the added degrees of freedom are not
accounted for.  To better assess the significance of the bimodal
structure we use the Akaike Information Criterion corrected for finite
sample size \citep[AICc;][]{burnham_model_2002}.  When comparing two
models, the one with the smallest AICc provides a better fit to the
data.  The AICc test is similar to a maximum likelihood ratio that
penalizes additional parameters.

The two models that we compare are the following: (1) The reference
model is a regular normal distribution, with two parameters -- a mean
and an intrinsic dispersion. (2) The alternative model is bimodal with
three parameters -- two means and the ratio of the two mode
amplitudes.  In the bimodal model no intrinsic dispersion is needed --
the SALT2 and peculiar velocity uncertainties are able to fully
explain the dispersion of each mode.  In both cases the integral of
the model is required to equal the number of SNe used in the fit.  For
the full sample we find $\Delta \text{AICc(bimodal - unimodal)} =
-5.6$, which strongly favors the bimodal structure.  If we remove the
possibly-contaminated ``p.m.s'' cases we find a slightly reduced value
of $\Delta \text{AICc(bimodal - unimodal)} = -4.3$.  Thus, whether we
include these or not has little impact.  The fact that the bimodal
model does not require the introduction of an \emph{ad~hoc} intrinsic
dispersion lends it further weight.

In addition we have tested for bimodality in the Hubble residuals for
the locally passive subset alone.  Here the first normal mode is
forced to match the mean \cHR{} of the \sS{} subset ($\mu_1=0.056$)
while the mean of second normal mode and the ratio of the two mode
amplitudes are free parameters.  Here again there is no need to
include intrinsic dispersion, thus there are only two parameters.
When comparing those two models, $\Delta \text{AICc(bimodal -
  unimodal)} = -3.5$, which again favors the bimodal structure, though
less strongly.

In Sect.~\ref{sec:mass-step} we will show how this peculiar structure
introduces the brightness offset observed between SNe in low- and
high-mass hosts.  This ``mass step'' is seen in our sample
\citep{childress_host_2012}, as well as in all other large surveys
\citep{kelly_hubble_2010, sullivan_dependence_2010,
  gupta_improved_2011, johansson_sne_2012}.  This suggests that this
bimodality is not exclusive to the SNfactory data, but is an intrinsic
property of SALT2-calibrated SNe~Ia.  That it has not been noticed
before is very likely due to the rarity of SNe~Ia having all of the
necessary ingredients -- well-measured lightcurves, small
uncertainties due to peculiar velocities and a wide range of
\SigmaHa{} and/or total stellar mass.

\section{Discussion}
\label{sec:Discuss}

In this section, we discuss the results previously presented.  First,
we look for the potential environmental dependency of stretch and
color coefficients among the two \SigmaHa{} subgroups
(Sect.~\ref{sec:standardization}).  Next, we investigate the
previously-observed relations between host masses and both the SNe
stretches (Sect.~\ref{sec:mass-stretch}) and the SNe corrected Hubble
residuals (Sect.~\ref{sec:mass-step}).  These relations are
interpreted from our local analysis perspective.

\subsection{The Standardization Process}
\label{sec:standardization}

In the previous section, the distributions of SN~Ia parameters have
been shown to vary between our two \SigmaHa{} subgroups.  In this
section, we investigate whether a unique standardization process
applies in the context of SALT2, by comparing stretch and color
coefficients ($\alpha{}$ and $\beta{}$ respectively; see
Eq.~\ref{eq:cHR}) between the \SigmaHa{} subgroups.

We find that the stretch coefficient, $\alpha$, is consistent between
\SigmaHa{} groups, having $\delta\alpha(\sS-\sN) = -0.020\pm0.037$.

The color coefficient of SNe between low-and high-\SigmaHa{} differ by
$\delta\beta(\sS-\sN) = 1.0 \pm 0.41$, but this difference is
dominated by the reddest SN, SNF20071015-000.  After removing it the
color coefficients are not significantly different:
$\delta\beta(\sS-\sN) = 0.67 \pm0.45$.  Therefore, our analysis is
insensitive to any color coefficient variations between \SigmaHa{}
groups.  The color of the outlier, SNF20071015-000, is likely
dominated by dust, for which a larger color coefficient is expected,
as we have shown in \citet{chotard_reddening_2011}.

We also measure what influence the lowest-stretch SNe~\sN{} from
passive environments may have on the existence of the brighter \Mtwo{}
mode.  Our sample contains 15 \sN{} with $x_1<-1$; we performed an
independent standardization on the remaining 67 SNe, among which 26
are \sN{}, 11 of which have $\cHR{}<-0.1$.  This fraction,
11~\Mtwo/26~\sN, is exactly the same as that computed from the full
sample (17/41).  Therefore, we conclude that the existence of the
\Mtwo{} mode is unrelated to the presence of a fast decline rate
subgroup in passive environments.

\subsection{Host Mass vs. SN properties: A Local Environment
  Perspective}
\label{sec:mass_SNe}

We investigate here the correlations found by global host studies,
namely between total host stellar mass and both the stretch and the SN
corrected Hubble residuals \citep{kelly_hubble_2010,
  sullivan_dependence_2010, gupta_improved_2011, childress_host_2012,
  johansson_sne_2012}.

\subsubsection{The Mass-Stretch Correlation}
\label{sec:mass-stretch}

\begin{figure}
  \centering
  \includegraphics[width=\linewidth]{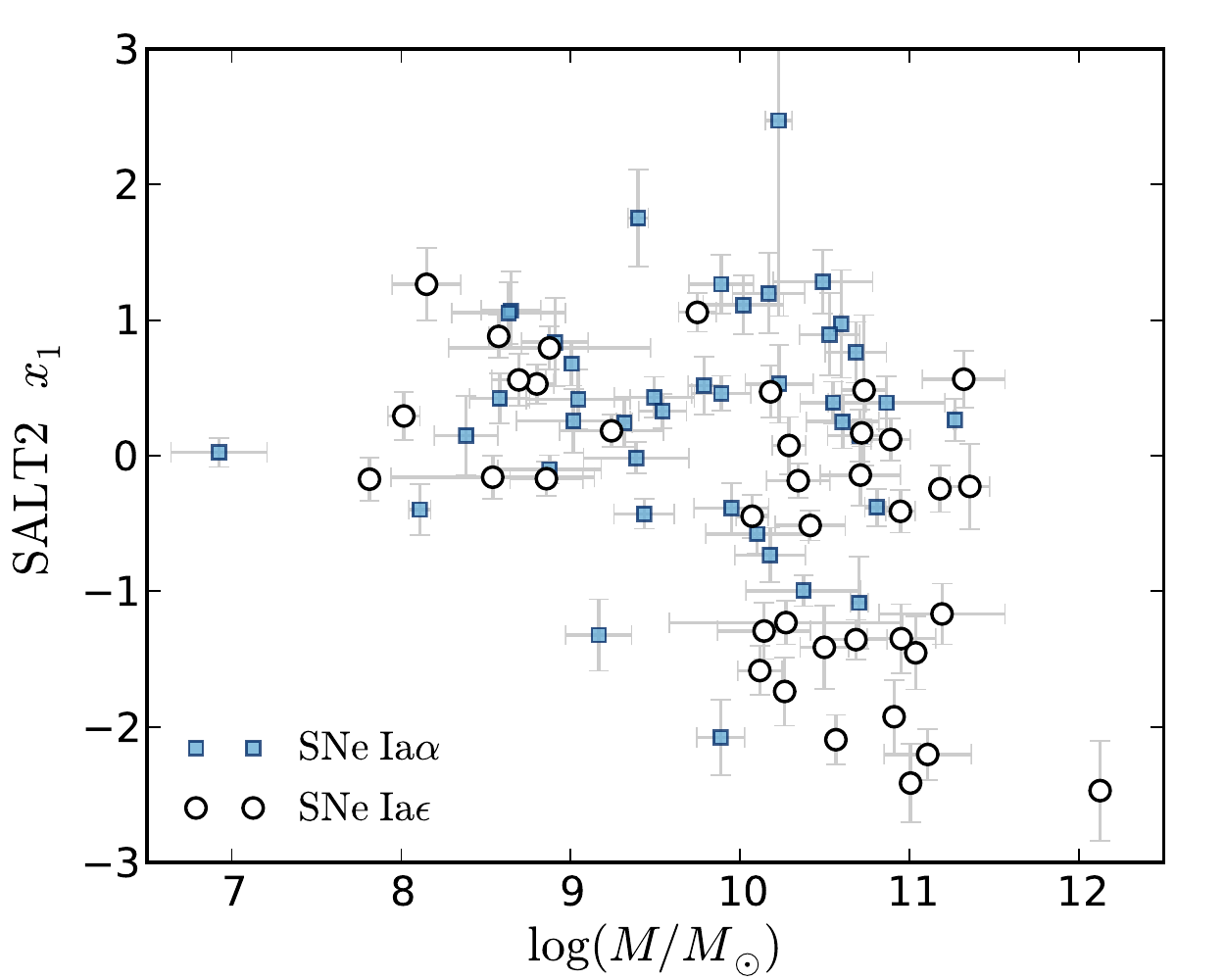}
  \caption{SN~Ia $x_{1}$ vs. total host stellar mass
    \citep{childress_host_2012}.  Markers follow the same color and
    shape code as in Fig.~\ref{fig:HaFrac__vs__salt2_HR}.  Note the
    distinctive ``L''-shaped distribution of the SNe~\sN.}
  \label{fig:Mass_X1}
\end{figure}

Stretch has been shown to correlate with host stellar mass
\citep[e.g.][]{sullivan_dependence_2010}.  Figure~\ref{fig:Mass_X1}
shows $x_1$ as a function of host mass for our sample \citep[see][for
details on the SNfactory host mass analysis]{childress_host_2012}.  In
agreement with the literature, we find a correlation between mass and
stretch when all the SNe~Ia are considered together, with a Spearman
rank correlation coefficient that is non-zero at the $3.4\sigma$ level
($r_s=-0.36$, $p_s=9.2\times 10^{-4}$).

When looking at \SigmaHa{} subgroups though, we notice that SNe~Ia
from locally star-forming environments (\sS) show no correlation
between their stretch and the mass of their host ($r_s = -0.02$ ;
$p_s=0.89$).  The original correlation is actually driven by SN~\sN{}
from locally passive environments ($r_s=-0.55$; $p_s=4.8\times
10^{-3}$, a $3.0\sigma$ significance).  More precisely, the
correlation appears to arise from the bimodal $x_1$ structure of this
subgroup, which we previously noted in
Fig.~\ref{fig:HaFrac__vs__salt2_HR}.  Figure~\ref{fig:Mass_X1} clearly
shows that the correlation is related to the L-shaped distribution of
SNe~\sN{} in the $x_1$--mass plane: SNe with $x_1>-1$ contribute to
the whole range of host masses, whereas the ones with $x_1<-1$ arise
exclusively in massive galaxies ($\hostmass > 10$).

From the local environment point-of-view, \emph{the light-curve widths
  of SNe in locally star-forming regions are independent of the total
  stellar mass of their hosts}: for SNe~\sN{}, the total host stellar
mass is not a stretch proxy as suggested by
\cite{sullivan_dependence_2010} but an indicator of whether the two
stretch subgroups ($x_1\lessgtr -1$) are represented or not.  From
this new perspective, the mass-stretch correlation shown Fig.~2 of
\cite{sullivan_dependence_2010} could be re-interpreted as being
driven by a relatively independent grouping of SNe having larger
masses ($\hostmass > 10.2$) and lower stretches \citep[$s<0.9 \sim
x_1<-1$;][]{guy_salt2:_2007}.

\subsubsection{The ``Mass Step''}
\label{sec:mass-step}

\begin{figure*}
  \centering
  \includegraphics[width=\textwidth]{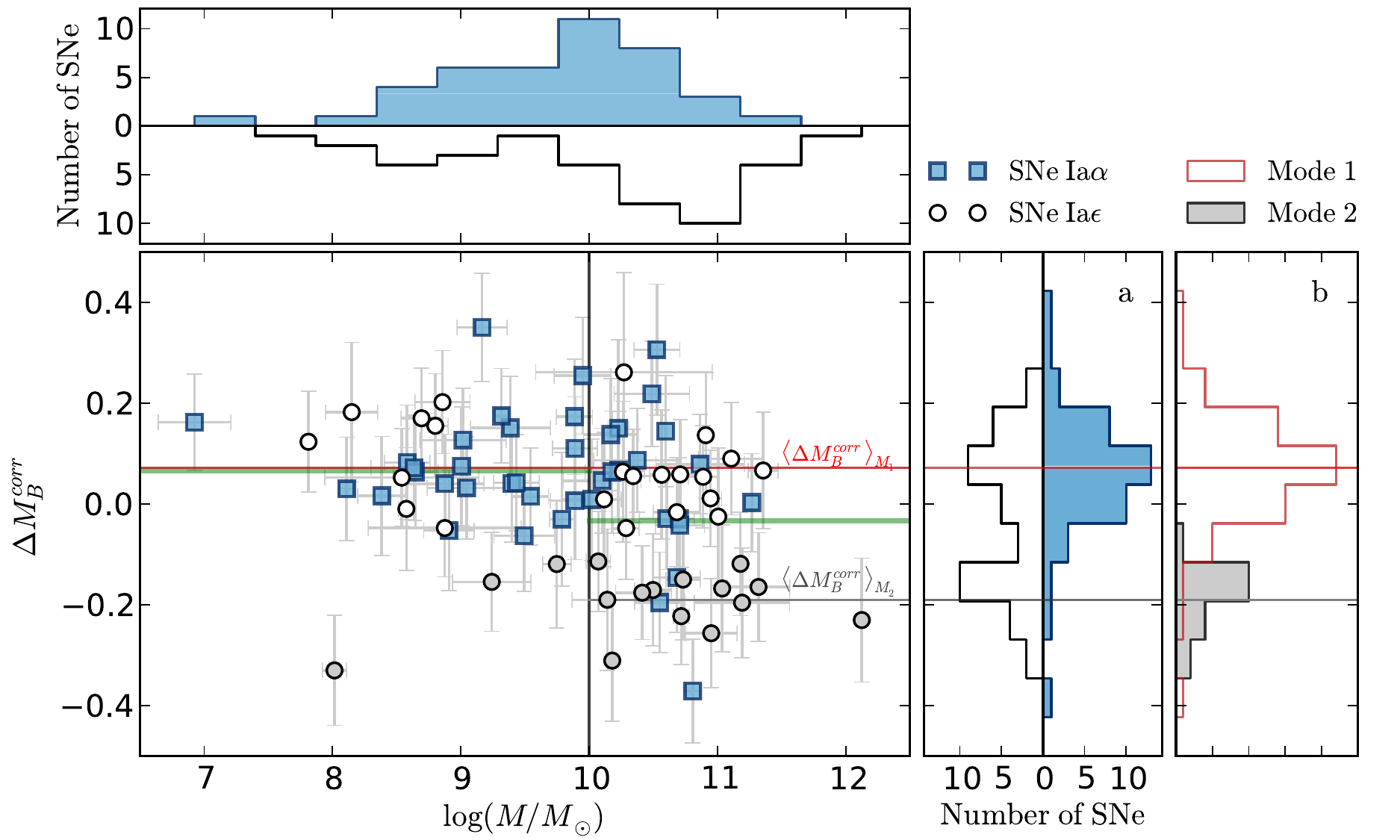}
  \caption{Corrected Hubble residuals as a function of the total host
    stellar mass, showing the mass step.  Markers follow the same
    color and shape code as in Fig.~\ref{fig:HaFrac__vs__salt2_HR}.
    \emph{Top:} total host stellar mass distribution per \SigmaHa{}
    subgroup.  \emph{Right:} \cHR{} distribution per \SigmaHa{}
    subgroup (\emph{a}) and per mode (\emph{b}), as shown in
    Fig.~\ref{fig:cHR__Two_modes}.  Horizontal red and gray lines
    indicate \Mone{} and \Mtwo{} weighted mean values respectively.
    \emph{Central panel:} Left and right green horizontal lines
    indicate the weighted mean \cHR{} of SNe~Ia hosted by galaxies
    more or less massive than $\hostmass = 10$ (vertical gray
    line). The difference between those two magnitudes defines the
    ``mass-step.''}
  \label{fig:mass-step}
\end{figure*}

The second known correlation between host masses and SNe~Ia properties
is a offset in corrected Hubble residuals between SNe hosted by
high-mass galaxies (brighter) and those from low-mass hosts (fainter).
This can be seen in Fig.~\ref{fig:mass-step}.  This mass effect has
been observed by all of the large SNe~Ia surveys
\citep{kelly_hubble_2010, sullivan_dependence_2010,
  gupta_improved_2011, childress_host_2012, johansson_sne_2012}, and
compiled in \cite{childress_host_2012}.  Hereafter, we will use the
term ``plateau'' to denote the averaged \cHR{} values for SNe from
either high ($H$) or low-mass ($L$) galaxies, with a dividing point
set at $\hostmass = 10$.  In \citet{childress_host_2012} we showed
that such a step function is a very good description of the data.  The
difference in the brightness of these two plateaus will be referred to
as the ``mass step.''  For our subset of the SNfactory sample this
offset reaches $\DcHRMS = -0.098\pm0.031$~mag.

We notice in Fig.~\ref{fig:mass-step} \citep[see also Fig.~2
of][]{childress_host_2012} that the upper part of the envelope of the
distribution is constant with mass whereas the lower part of the
distribution extends to bright \cHR{} values for masses above
$\hostmass \sim 10$. We see that the high-mass plateau is brighter
only because a subgroup of bright SNe exists in massive hosts.  A
similar trend can be observed in Fig.~4 of
\cite{sullivan_dependence_2010} for SNLS, in Fig.~15 of
\cite{johansson_sne_2012} for SDSS, and in Fig.~5 of
\cite{childress_host_2012} where we combined all available data.

We see in the top panel of Fig.~\ref{fig:mass-step} that the local
\Ha{}~environment changes with the total host stellar mass.  For low
masses up to $\hostmass \la 10$, locally star-forming environments
dominate (23 \sS/34).  In more massive hosts ($\hostmass \ga 10$),
however, the local SN environment presents a marked change and the
locally passive cases become favored (27 \sN/45).  Since only a few
\Mtwo{} SNe~\sN{} are present in lower mass hosts (3 \Mtwo{}/34), the
first plateau happens to be equal to the mean \Mone{} value, with
$\langle\cHR\rangle_{L} = 0.066 \pm 0.021$.  In the upper mass range,
because of the rise of the \sN{} subgroup populating the brighter
\Mtwo{} mode, the average brightness increases up to the level of the
second plateau. The \sS{} population shows no appreciable mass step;
we find $\delta \langle M_B^{\mathrm{corr}}\rangle_{\sS} (H-L)
=-0.06\pm0.04$, largely driven by the bright SNF20070701-005.  There
is a strong possibility that this SN~Ia is a false positive \Ha{}
association (see Sect.~\ref{sec:SNeIS}), and if we remove it we find
$\delta \langle M_B^{\mathrm{corr}}\rangle_{\sS} (H-L) =-0.04\pm0.04$.
\emph{We conclude therefore that the existence of a brighter mode,
  completely dominated by SNe having locally passive environments,
  causes both the \Ha{}~bias and the mass step.}

\section{Consequences for Cosmology}
\label{sec:Cosmo_consequences}

The proportion of SNe from locally passive environments has surely
changed with time, so the fraction of \Mtwo{} SNe will be
redshift-dependent, and thus we expect the amplitude of the mass step
will change as well.  In this section we develop a first estimate of
this evolution bias and its consequence for measurements of
cosmological parameters.  In Sect.~\ref{sec:influence_w}, we predict
the redshift dependence of the magnitude bias due to evolution in the
relative numbers of SNe~\sS{} and SNe~\sN{} given the \Ha{}~bias we
observe. Using this, we then estimate the bias on measurement of the
dark energy equation of state parameter, $w$, that would result if
this evolution does indeed occur. By then coupling the \Ha{}~bias and
the mass step as discussed in Sect.~\ref{sec:ms_model}, we verify the
predicted evolution of the mass step using data from the literature.
In Sect.~\ref{sec:SNeIS}, we introduce a subclass of SNe whose
Hubble-residual dispersion is significantly reduced by avoiding this
bimodality.  We discuss possible strategies for mitigating this bias
in Sect.~\ref{sec:mitigation}.

\subsection{Evolution of the SNe~Ia Magnitude with Redshift}
\label{sec:influence_w}

\subsubsection{Evolution Model}
\label{sec:toy_model}

The mean sSFR is known to increase with redshift
\citep[e.g.][]{perez-gonzalez_stellar_2008}, which most likely
decreases the proportion of SNe associated with locally passive
environments \citep{sullivan_rates_2006, hopkins_normalization_2006}.
Because of the \Ha{}~bias, the average SN brightness is expected to be
redshift dependent.

We define $\psi(z)$ as the fraction of SNe located in locally passive
environments (\sN) as a function of redshift.  Assuming that the
brightness offset between \sS{} and \sN{} is constant with redshift,
the mean corrected $B$-magnitude of SNe~Ia at maximum light can then
be written as:
\begin{align}
  \label{eq:cH_z}
  \langle\cH\rangle (z) &=
  (1-\psi(z)) \times \langle\cH\rangle_{\sS} +
  \psi(z) \times \langle\cH\rangle_{\sN} \nonumber\\
  &= \langle\cH\rangle(\sS) + \psi(z)\times \DcHR.
\end{align}

\cite{perez-gonzalez_stellar_2008} demonstrate that the global
galactic sSFR increases by a factor of $\sim 40$ between $z=0$ and~$2$
\citep[see also][]{damen_evolution_2009}.  This evolution can be
adequately approximated as:
\begin{equation}
  \label{eq:sSFR_Z}
  \log(\mathrm{sSFR})(z) \simeq \log(\mathrm{sSFR}_{0}) + 0.95 \times z.
\end{equation}
Since sSFR measures the fraction of young stars in a galaxy, we
suggest as a toy model that the proportion of SNe from locally passive
environments evolves in complement \citep{sullivan_rates_2006}
and therefore decreases with $z$:
\begin{equation}
  \label{eq:psi_z_def}
  \psi(z) = (K \times 10^{0.95z} + 1)^{-1}.
\end{equation}
By definition of the \sS/\sN{} subgroups, the SNfactory sample sets
the normalization to $\psi(z=0.05) = 41/82 = 50\pm5\%$
\citep{cameron_estimation_2011}; hence $K=0.90 \pm 0.15$.

Then, using SN, CMB, BAO, $H_0$ constraints as in
\cite{suzuki_hubble_2012}, along with the assumption that the Universe
is flat, we estimate the impact of a bias in SN Ia luminosity distance
measurements.  Relative to existing SN cosmological fits, we find that
shifting the SN Ia luminosity distances according Eqs.~\ref{eq:cH_z}
and \ref{eq:psi_z_def} would shift the estimate of the dark energy
equation of state by $\Delta w \approx -0.06$ (D. Rubin, private
communication, as estimated using the Union machinery,
\citealt{kowalski_improved_2008, amanullah_spectra_2010,
  suzuki_hubble_2012})

\subsubsection{Prediction of the Mass Step Evolution}
\label{sec:ms_model}

The \Ha{}~bias is due to a brightness offset between SNe~Ia from
locally star-forming and locally passive environments.  Guided by
Fig.~\ref{fig:mass-step}, in Sect~\ref{sec:mass-step} we then made a
connection between the \Ha{}~bias and the mass-step by noting the
strong change in the proportion of SNe~\sS{} and SNe~\sN{} on either
side of the canonical mass division at $\hostmass = 10$.

In Appendix~\ref{sec:bias_math} the mathematical connection between
the \Ha{}~bias and the amplitude of the mass step is laid out. To a
good approximation, the near absence of bright SNe~\sN{} among
low-mass hosts means that the low-mass plateau is equal to the average
magnitude of the SNe~\sS{}, and is therefore constant with redshift.
By contrast, the brighter high-mass plateau has been shown to arise
from a subgroup of bright SNe~\sN; thus, the amplitude of the
mass-step will evolve along with the fraction of SNe~\sN{}, as given
by $\psi(z)$.

Thus, while local \Ha{} measurements are not available for
high-redshift SNe~Ia in the literature, our hypothesis -- that the
redshift evolution of the fraction of SNe~\sN{} given our toy model
Eq.~\ref{eq:psi_z_def} induces a reduction of the influence of the
bright SNe that create both the mass-step and the \Ha{} bias -- can be
tested using the same evolutionary form, $\psi(z)$.  We consequently
predict (see Appendix~\ref{sec:bias_math_obs}) that the amplitude of
the mass-step, \DcHRMS, evolves as:
\begin{equation}
  \label{eq:mass-step_z}
  \delta \langle M_B^{\mathrm{corr}}\rangle(H-L; z) \simeq \psi(z) \times
  \frac{\delta \langle M_B^{\mathrm{corr}}\rangle(H-L;
    z=0.05)}{\psi(z=0.05)},
\end{equation}
where setting the constant term, $\delta \langle
M_B^{\mathrm{corr}}\rangle(H-L; z=0.05)$, equal to our measured mass
step provides the normalization.

\subsubsection{Verification of Mass-Step Evolution}

\begin{figure}
  \centering
  \includegraphics[width=\linewidth]{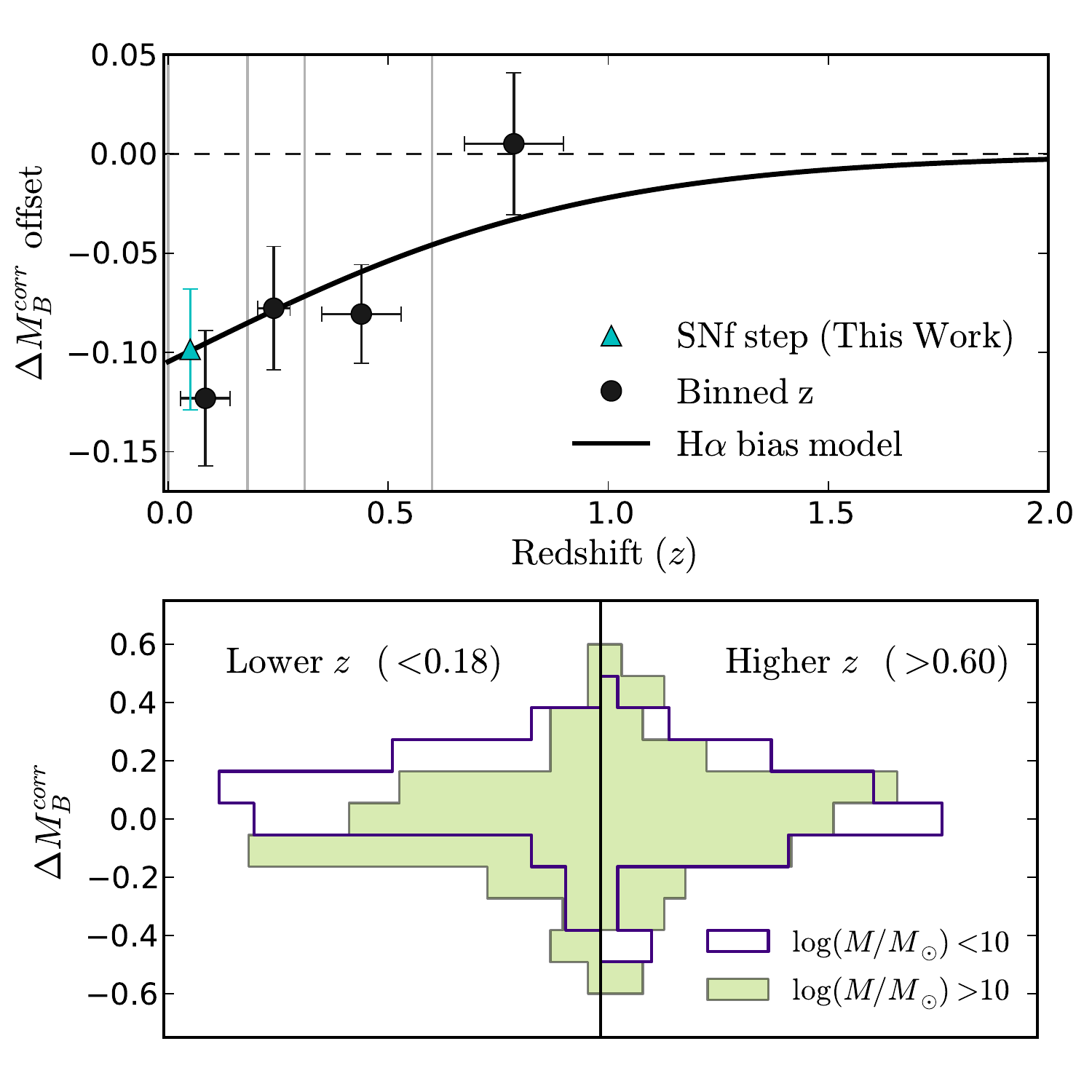}
  \caption{Observed and predicted redshift evolution of the \cHR{}
    offset between SNe from low- and high-mass hosts ($\hostmass
    \lessgtr 10$), i.e the ``mass step''.  \emph{Top:} black markers
    show the observed mass step offset per redshift bin (delimited by
    gray vertical lines).  Error bars in redshift represent the
    dispersion within each redshift bin.  The black line is the
    prediction from Eq.~\ref{eq:psi_z_def}, whose amplitude has been
    set to the SNfactory mass step amplitude at $z=0.05$ (filled-cyan
    triangle).  \emph{Bottom:} Distribution of the \cHR{} of SNe~Ia
    from low- (filled-yellow) and high-mass (open-purple) hosts.  The
    left histograms correspond to the closest SNe ($z<0.18$) while the
    right ones are for the highest-$z$ SNe ($z>0.6$).  The difference
    within a redshift bin of the mean value for low- and high-mass
    distributions defines the mass step at this redshift, as plotted
    in the upper panel.}
  \label{fig:Cosmo_plot}
\end{figure}

Figure~\ref{fig:Cosmo_plot} shows the predicted redshift evolution of
the amplitude of the mass step assuming our simple
SNfactory-normalized model. We compare this model to mass steps
measured using literature SNe from SNLS, SDSS and non-SNfactory
low-$z$ datasets \citep[][respectively]{sullivan_dependence_2010,
  gupta_improved_2011, kelly_hubble_2010} from the combined dataset of
\cite{childress_host_2012}.  We split those data into four redshift
bins with equal numbers of SNe~Ia ($\approx 120$ SNe~Ia in $z<0.18$, $
0.18<z<0.31$, $0.31<z<0.6$ and $0.6<z$ ranges), and we measure for
each bin the magnitude offset between SNe in low- and high-mass hosts
for a step located at $\hostmass=10$.  Figure~\ref{fig:Cosmo_plot}
shows that our simple model qualitatively reproduces the measured mass
step evolution with redshift. Relative to a fixed mass-step anchored
by the SNfactory data, our model gives $\Delta\chi^2 = -5.7$ for the
literature data sets. This provides external confirmation of behavior
consistent with the \Ha{}~bias at greater than 98\% confidence.

\subsubsection{Interpretation and Literature Corrections}

From this observation we draw several conclusions.  (1) The amplitude
of the mass step indeed decreases at higher redshift, so host mass
cannot be used, as it has been, as a third SN standardization
parameter \citep{lampeitl_effect_2010, sullivan_dependence_2010}.  (2)
The observed mass steps follow the predicted evolution based on the
\Ha{}~bias quite well.  This in turn lends further support to the idea
that the \Ha{}~bias is the origin of the magnitude offset with host
mass.  As the mass step bias is observed in different data sets, the
\Ha{}~bias appears to be a fundamental property of SNe~Ia standardized
using SALT2.  (3) Any \emph{ad~hoc} correction of the mass step by
letting SNe~Ia from low- and high-mass hosts have different absolute
magnitudes will be inappropriate as this assumes that high-mass hosted
SNe are brighter in a constant way \citep{conley_supernova_2011},
ignoring the observed redshift evolution.  Doing so will bias the
SN~\sS{} population from high-mass hosts that dominate at higher $z$,
and miss the observed evolution of mean SNe~Ia magnitude.  Therefore,
such a correction will not remove the bias on the dark energy equation
of state, which we estimate to be $\Delta w \sim 0.06$.

Of course application of a fixed mass-step correction remains
useful for bringing into agreement two datasets at the same redshift
that have different proportions of low-mass and high-mass host
galaxies due to survey selection effects.  Whether or not a
correction using a fixed mass-step might accidentally help or hurt
in correcting the redshift evolution of the \Ha{}~bias depends on
the -- currently unknown -- correlation of the parameters of
Eqs.~\ref{eq:cH_z} and \ref{eq:psi_z_def} with the redshift
evolution in the ratio of high- to low-mass host galaxies.

\subsection{Type Ia$\alpha$ SNe: Homogeneous SALT2 Candles}
\label{sec:SNeIS}

\begin{figure}
  \centering
  \includegraphics[width=0.8\linewidth]{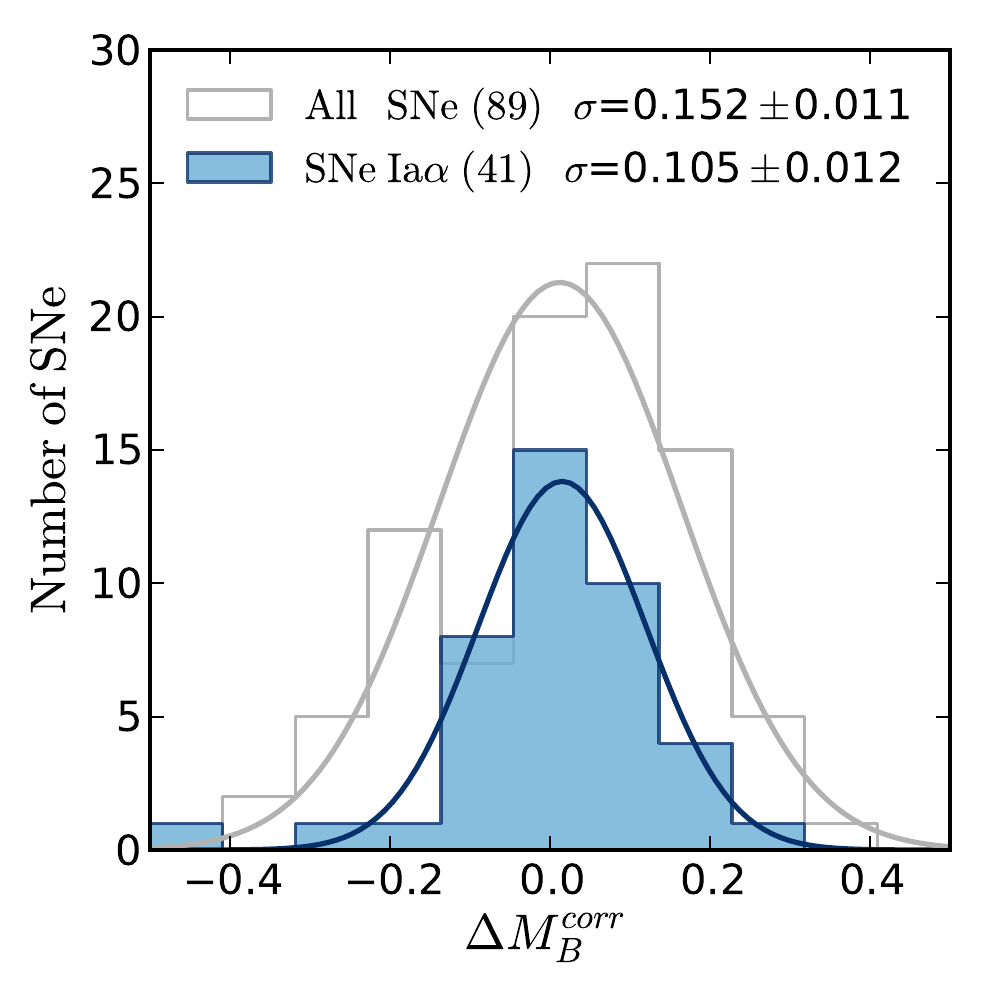}
  \caption{SN~\sS{} corrected Hubble residual distributions
    (\cHR(\sS), blue-filled histogram), and its best Gaussian fit
    (solid-blue line, one $3\sigma$-outlier rejected).  The
    standardization has been performed on SNe~\sS{} only.  For
    comparison, the full sample residual distribution (for a full
    sample standardization) is shown by the open histogram, with its
    best Gaussian fit in gray.  Both distributions where generated
    from independent standardizations and are therefore centered on
    zero.}
  \label{fig:cHR_IS_standardization}
\end{figure}

The SNe~\sS{} are unimodal in stretch and in corrected Hubble
residuals, and are therefore free from the aforementioned \Ha{}
bias/mass step.  The \cHR{} distribution for this new group of
supernovae, when the SALT2 standardization is performed on this
subsample only, is shown in Fig.~\ref{fig:cHR_IS_standardization}.

SNe~\sS{} standardization generates a smooth symmetric residual
distribution, with a significantly smaller magnitude dispersion of
$\sigma=0.105 \pm 0.012$~mag (after $3\sigma$ clipping) than the full
initial sample ($\sigma=0.152 \pm 0.011$~mag, no outlier).  This
dispersion is in agreement with the original \Mone{} fit ($0.099 \pm
0.009$~mag), and corresponds to a significant reduction of $\sim30\%$
of the Hubble residual dispersion.

In order to check the robustness of this result we conducted several
additional tests.  First we considered the effect of including the
$3\sigma$ outlier, SNF20070701-005.  Doing so increases the residual
standard deviation to $0.126 \pm 0.014$~mag (nMAD of 0.108~mag), but
we note that this outlier falls within the brighter part of the
distribution, potentially polluted by \Mtwo{} SNe; this SN might just
be a case of misassociation with a star-forming environment, as
discussed in Sect.~\ref{sec:Analysis}.  This possibility is reinforced
by the fact that this host is globally passive, having
$\log(\mathrm{sSFR}) = -10.8$ \citep{childress_host_2012}.

Next, we performed a permutation test based on independent
standardization trials for 41~randomly selected SNe from the full
sample.  The results attest to the high significance of the
environmental selection, giving $p=1.9\times 10^{-3}$.

Finally, $K$-fold cross-validation was employed in order to measure
the predictive power of our model.  Since it avoids over-fitting the
sample, cross-validation results are not expected to be as good as the
original fit \citep{kim_standardizing_2013}.  We find $\sigma=0.117
\pm 0.014$~mag using $k=10$, compatible with our original results.
The quoted error includes a marginal dispersion of 0.004~mag from
random ``fold'' selections.

\subsection{Mitigation Strategies}
\label{sec:mitigation}

The \Ha{}~bias is a concern for cosmology, as it reveals a local
environmental dependency -- most probably a progenitor effect -- not
accounted for by current SN standardization methods.  It is necessary
for ongoing and future surveys to consider it.  The best solution
would be if additional light curve or spectral features could be
identified that can be used to remove the impact of local host
properties.  Another possibility would be to include the bimodality in
the cosmology fitter, treating the redshift dependence and relative
fraction of passive environments as free parameters.  However, this
seems perilous as it could mask or induce signatures of exotic dark
energy, and likely would require lightcurves with much better
statistical accuracy in order to resolve the bimodality.
Implementation of our model from Sect.~\ref{sec:toy_model} also
  could be considered.  Examining these and other model-dependent
  correction possibilities is left for future work.  Alternatively,
it could prove useful to focus on the \sS{} subgroup for cosmological
analyses, since these appear to be free of environmental biases.

For cosmological measurements naturally it is not permissible to split
the SN~Ia sample based solely on Hubble diagram residuals, e.g. one
cannot directly select the homogeneous Mode~1 SNe only.  However, one
may select a subsample of SNe~Ia based on their local environment
properties: it is then possible to remove the brighter Mode~2 --
exclusive to locally passive environments -- by discarding the entire
\sN{} group.  This is not as draconian as it may sound; while only
$\sim50$\% of the sample would be retained, this subset has roughly
$1.8\times$ as much statistical weight per SN.  Thus, the statistical
power is nearly the same as using the full sample.  Of course
obtaining local environment measurements could be observationally
expensive, but Fig.~\ref{fig:Cosmo_plot} indicates that most of the
bias is at lower redshifts, where local environment measurements are
less difficult.  Since \sS{} appear to be dustier, this might add some
complications, but in our dataset the \sS{} standardize well despite
this.  We note that removing SNe from globally passive hosts will only
remove approximately half of the SNe~\sN{}.  However, at least then local
measurements would not be needed for these.

In conclusion, SNe~Ia from passive environments (\sN) appear to cause
the known (yet not understood) environmental biases, and we raise the
possibility of removing them for cosmological studies.  As SNe~\sS{}
form an homogeneous subgroup of standardizable candles, free from
environmental biases and therefore showing a tighter dispersion on the
Hubble diagram, they could be used exclusively for cosmological
analyses.

\section{Robustness of the Analysis}
\label{sec:Assumption_discussion}

In this section, we analyze the robustness of our results, by testing
potential issues in our analysis: (1) the impact of seeing given our
1~kpc aperture; (2) the influence of SNe potentially misassociated
with \Ha{} signal from host galaxy core; (3) the influence of the
\SigmaHa{} boundary defining the \sN{} and \sS{} subgroups.  The
general conclusion of these tests is that our results are not
significantly affected by changes in our assumptions.

\subsection{Restframe Aperture and Redshift Bias}
\label{sec:issue_PFS}

All our spectra have been averaged over an identical spatial aperture
of 1~kpc radius (see Sect.~\ref{sec:merge-cubes}).  However, the
atmospheric seeing PSF correlates spaxels on the arcsecond scale.
For the SNfactory dataset the median seeing is 1\farcs1.  This
corresponds to a projected correlation length of $\sim$0.65~kpc at
$z=0.03$; $\sim$1.1~kpc at $z=0.05$, and $\sim$1.6~kpc at $z=0.08$.
The redshift therefore has an influence on the amount of independent
information in the 1~kpc radius aperture around the SN due to the
seeing.  Seeing also allows contamination of the local environment
from bright sources outside the aperture.

However, a Spearman rank correlation analysis shows no correlation
between $\log(\SigmaHa{})$ and $z$ ($r_s=0.17$, $p_s=0.14$), and both
\SigmaHa{} subgroups have similar redshift distributions
($p_{KS}=0.25$).  Consequently, we believe that our analysis is
relative unaffected by a redshift bias induced by seeing.

\subsection{SNe with Potentially Misassociated \Ha{} signal}
\label{sec:issue_poe}

SNe close to the center of very inclined hosts (referred to as
``p.m.s'', see Sect.~\ref{sec:Analysis}) have been directly removed
from the main analysis, given the possible misassociation with \Ha{}
from the core of the galaxy.  However, when we include those SNe, none
of the previous results are significantly impacted:
\begin{description}
\item[Color Relation:] The observed correlation between $c$ and
  \SigmaHa{} is slightly more significant, with a Spearman rank
  correlation coefficient of $r_s=0.22$ ($p_s=0.035$);
\item[\Ha{}~bias:] The bias increases marginally, with $\DcHR = -0.102
  \pm 0.031$~mag, ($p_{KS}=6.5\times 10^{-3}$);
\item[Stretch and host mass:] $x_{1}$ and the total host stellar mass
  remain uncorrelated for SNe~\sS{} ($r_s=0.07$, $p_s=0.66$);
\item[SNe~\sS{} dispersion:] The \sS{} \cHR{} dispersion increases to
  $0.115 \pm 0.012$~mag ($p=0.011$), which is still compatible with
  our fiducial result.
\end{description}

\subsection{Influence of the \sS/\sN{} Boundary}
\label{sec:issue_limi}

In our fiducial analysis the division between locally passive (\sN{})
and locally star-forming (\sS) subgroups was set at the median value
$\log(\SigmaHa) = 38.35$ of a 82~SN sample.  The confidence interval
for this population split is 37--45 \citep{cameron_estimation_2011},
and we now demonstrate the robustness of our analysis to these
\sS/\sN{} limit shifts.
\begin{description}
\item[\Ha{}~bias:] The bias changes within the errors bars, with
  $\DcHR = -0.101 \pm 0.031$~mag and $\DcHR =-0.075 \pm 0.032$~mag for
  37 and 45 \sN, respectively.
\item[Stretch and host mass:] $x_1$ and the total host stellar mass of
  the SNe~\sS{} remain uncorrelated in both cases, giving $p_s=0.84$;
\item[SNe~\sS{} dispersion:] The \sS{} \cHR{} dispersion is the same
  within the error bars, with $0.115 \pm 0.012$~mag and $0.112 \pm
  0.012$~mag for 37 and 45 \sN, respectively.
\end{description}

Further, in Sect.~\ref{sec:Analysis} we pointed out cases of SNe~\sN{}
in globally star-forming hosts located near our division point, e.g.,
having $\log(\SigmaHa)\sim38$ and $\mathrm{sSFR}>-10$.  Whether or not
they truly are locally passive or locally star-forming is therefore
somewhat uncertain.  Here we show that our main results %
do not depend on whether or not those SNe are included in the \sS{}
subgroup.  However, three of these cases are bright and have been
counted as \Mtwo{} in our fiducial analysis, therefore moving them from
\sN{} to \sS{} increases the dispersion of Hubble residuals of the
SNe~\sS{} subset.

\begin{description}
\item[\Ha{}~bias:] The bias marginally changes, with $\DcHR = -0.097
  \pm 0.032$~mag,
\item[Stretch and host mass:] $x_1$ and the total host stellar mass of
  the SNe~\sS{} remain uncorrelated, giving $p_s=0.57$;
\item[SNe~\sS{} dispersion:] The \sS{} \cHR{} dispersion is degraded
  by the inclusion of bright SNe that skew the normal distribution:
  $\sigma = 0.144 \pm 0.014$~mag (no SNe have been clipped).
\end{description}

\section{Summary \& Conclusion}
\label{sec:Conclusion}

Using the SuperNova Integral Field Spectrograph (SNIFS), we have
assembled a sample of 89~nearby SNe~Ia in the redshift range
$0.03<z<0.08$ with accurate measurements of the \Ha{} surface
brightness within a 1~kpc radius (\SigmaHa) used as a tracer of young
stars in the SN neighborhood.

This sample has been split in two at the median value
$\log{\SigmaHa{}}=38.35$, which happens to correspond to the mean warm
interstellar medium \Ha{} surface brightness of galaxies
\citep{oey_survey_2007} and to the mean \Ha{} sensitivity level at our
highest redshift.  SNe with low \SigmaHa{} have then been referred to
as ``locally passive'' (\sN) while those with high \SigmaHa{} are
designated as ``locally star-forming'' (\sS).  The wide range of
measured \SigmaHa{} indirectly confirms the existence of SNe~Ia
associated to both young and old stellar populations.  We have then
studied variations of SNe properties as a function of this local host
parameter.

\paragraph{Color:}
Our local analysis highlights a $2\sigma$ significance correlation
between $\log(\SigmaHa{})$ and the SALT2 color $c$, with the SNe~\sS{}
being redder, as expected \citep{charlot_simple_2000}.

\paragraph{Stretch:}
The relationship between SN light-curve stretch and local star
formation activity is more complex than that suggested by previous
global host studies.  Faster declining SNe ($x_{1} < -1$) are indeed
strongly favored by locally passive environments, but passive local
environments also host SNe with moderate stretches.  Lower stretch
SNe~Ia are hosted in locally passive environments, but we find that
moderate stretch SNe~Ia are hosted by all types of environments.  The
local perspective also shows that the correlation between the host
stellar mass and SN stretch is entirely created by the SNe~\sN{}.  The
stretches of SNe~\sS{}, from locally star-forming regions are
uncorrelated with the total stellar masses of their hosts.

\paragraph{Hubble residuals:}
After routine SALT2 standardization using stretch and color, we have
shown that the SN~Ia Hubble residuals (\cHR) are significantly
environment-dependent.  SNe~\sN{} from locally passive regions are
$0.094 \pm 0.031$~mag brighter than SNe~\sS{} from locally
star-forming regions.  This \Ha{}~bias originates from a bimodal
structure in the Hubble residual distribution.  The first mode,
denoted as \Mone{}, is present in all environments, independent of the
local star formation activity.  The second mode, \Mtwo{}, is
intrinsically brighter and exclusive to locally passive environments,
and predominatly occurs in high-mass hosts.  We noted that this
peculiar structure naturally explains the previously observed ``mass
step'', i.e. the offset in brightness between SNe from low- and
high-mass hosts \citep[e.g.][]{childress_host_2012}.  This suggests
that the \Ha{} bias is not exclusive to our sample, and is of a more
fundamental origin than the global host effect.  Additional SN data,
with measurement uncertainties below the 0.1 mag, will be critical for
confirming the Hubble residual bimodality.  Our analysis does not
suggest any significant dependence on the local \Ha{} environment of
the stretch or color coefficients used in standardization.

\paragraph{Cosmological implications:}
The \Ha{}~bias is a concern for cosmology since the fraction of SNe in
star-forming environments most likely changes with redshift, which
implies an evolution with $z$ of the mean $B$-magnitude of SNe~Ia
after standardization.  Using a simple toy model based on the
evolution of the specific star formation rate of galaxies with
redshift, we estimated this effect could shift the measurement of the
dark energy equation of state parameter by $\Delta w\sim
-0.06$. Currently-employed constant mass-step corrections might
accidently reduce or increase the bias on $w$, but do not remove it.
Moreover, the amplitude of the mass step is shown to evolve with
redshift as predicted, which supports our claim that both the
\Ha{}~bias and the mass-step arise from the same underlying
phenomenon.  Future cosmological analyses will need to consider these
environmental dependencies. An ideal solution would be to find
additional light curve or spectral features able to correct this
bias. Another solution would be to focus on a subsample of SNe~Ia free
from this known --~yet not understood~-- biases.

\paragraph{SNe~\sS{} as homogeneous SALT2 candles:}
SNe~Ia from locally passive environments (\sN) appear to be the cause
of all the biases we have observed.  We have raised the idea of
removing them from cosmological fits for this reason.  The remaining
bias-free SNe~\sS{} have an intrinsic Hubble residual dispersion
reduced to $0.105\pm0.012$~mag, suggesting that a selection on local
environment would significantly improve the quality of SNe~Ia as
cosmological distance indicators.  We further noted that this
remaining dispersion is entirely consistent with our measurement
uncertainties; no intrinsic dispersion is necessary.  The trade-off
would be minor loss in net statistical power compared to using the
full SN sample and the need for measurements of the local environment,
versus a far more serious bias.

At this juncture SN cosmology is limited by systematic errors.
\citep{conley_supernova_2011, clair_paper_2013}.  Calibration issues
are thought to set the current limits, but our first analysis of the
local SN environments has revealed strong correlations between SNe and
the star-forming activity of their immediate environment.  This is
most likely caused by our limited knowledge of the progenitors of the
SNe~Ia, the existence of potential SN subgroups, and the evolution
with redshift of these standardizable candles.

For future SN cosmology efforts we thus suggest further studies
exploring the properties of SN~Ia local host environments.  Notably,
the following local host properties may bring essential clues leading
to the minimization of progenitor dependencies: gas and stellar
metallicities; star formation history surrounding the SNe, and local
interstellar dust extinction.  Continuing to refine these measurements
will constitute a major step forward in the direction of constructing
an accurate ``likes-to-likes'' SN cosmology.

\begin{acknowledgements}
  We thank Dan Birchall for observing assistance, the technical and
  scientific staffs of the Palomar Observatory, the High Performance
  Wireless Radio Network (HPWREN), and the University of Hawaii 2.2~m
  telescope.  We recognize the significant cultural role of Mauna Kea
  within the indigenous Hawaiian community, and we appreciate the
  opportunity to conduct observations from this revered site.  This
  research was conducted within the framework of the Lyon Institute of
  Origins under grant ANR-10-LABX-66. M.R. thanks French Rhône-Alpes
  region for support from its grant Explo'ra doc 11-015443-03.  This
  work was supported by the Director, Office of Science, Office of
  High Energy Physics, of the U.S. Department of Energy under Contract
  No. DE-AC02-05CH11231; by a grant from the Gordon \& Betty Moore
  Foundation; in France by support from CNRS/IN2P3, CNRS/INSU, and
  PNCG; and in Germany by the DFG through TRR33 ``The Dark Universe.''
  Some results were obtained using resources and support from the
  National Energy Research Scientific Computing Center, supported by
  the Director, Office of Science, Office of Advanced Scientific
  Computing Research, of the U.S. Department of Energy under Contract
  No. DE-AC02-05CH11231.  HPWREN is funded by National Science
  Foundation Grant Number ANI-0087344, and the University of
  California, San Diego.  Based in part on observations made with the
  NASA Galaxy Evolution Explorer, operated for NASA by the California
  Institute of Technology under NASA contract NAS5-98034, with
  analysis supported by GALEX Archival Research Grant
  \#08-GALEX508-0008 for program GI5-047 (PI: Aldering).
\end{acknowledgements}

\bibliographystyle{aa}
\bibliography{bibliography}

\begin{appendix}
  \newpage

  \onecolumn

  \section{Relation between the \Ha{} bias and the Mass Step}
  \label{sec:bias_math}

  In this Appendix we provide the mathematical details needed to
  understand the relation between the \SigmaHa{} group magnitudes and
  the observed mass step. This is all that is needed to understand the
  relation between the numerical values of the \Ha{} bias and the mass
  step in the SNfactory dataset.

  To begin, the dataset is partioned into four subsets consisting of
  SNe~\sS{} and SNe~\sN{} above and below the canonical mass division
  point of $\hostmass = 10$. We denote the number in each subset as
  $N_{\alpha,L}$, $N_{\alpha,H}$, $N_{\epsilon,L}$ and
  $N_{\epsilon,H}$, with $L$ and $H$ symbolizing SNe~Ia in low- and
  high-mass hosts, respectively.  Likewise, the mean Hubble residuals
  for each subset are denoted as $\langle\cHR\rangle_{\alpha,L}$,
  $\langle\cHR\rangle_{\alpha,H}$, $\langle\cHR\rangle_{\epsilon,L}$
  and $\langle\cHR\rangle_{\epsilon,H}$.

  In order to make the equations more readable we introduce the
  following notation for fractions: $F^{A}_{B}$ means ``the fraction
  of SNe from subset $A$ that have property $B$''.  For instance
  $F^{H}_{\epsilon}$ means ``the fraction of SNe~Ia from high-mass
  hosts that are \sN{}'', i.e., $N_{\epsilon,H}/N_H$.  Mathematically,
  $B$ is the numerator and $A$ is the denominator.  Naturally, the
  partitioning imposes the requirement that $F^{A}_{B} +
  F^{A}_{\neg{B}} = 1$, e.g., $F^{H}_{\epsilon} + F^{H}_{\alpha} =
  (N_{\epsilon,H}+N_{\alpha,H})/N_{H} = 1$.

  \subsection{General equations}
  \label{annex:general}

  The \Ha{} bias, \DcHR, is defined as the weighted mean difference of
  the \cHR{} between the \sN{} and \sS{} subsamples.
  \begin{equation}
    \label{annex:eq:ha-def}
    \DcHR \equiv \langle\cHR\rangle_{\sN} - \langle\cHR\rangle_{\sS}
  \end{equation}
  The mass-step, \DcHRMS{}, is defined as the weighted mean difference
  of the \cHR{} between SNe~Ia in low-mass and the high-mass hosts.
  \begin{equation}
    \label{annex:eq:mass-step-def}
    \DcHRMS \equiv \langle\cHR\rangle_{H} -\langle\cHR\rangle_{L}
  \end{equation}
  The individual weighted mean magnitudes entering
  Eqs.~\ref{annex:eq:ha-def} and \ref{annex:eq:mass-step-def} can be
  written in terms of the weighted mean magnitudes for each
  portion. Introduction of our $F^{A}_{B}$ notation then gives
  \begin{align}
    \label{eq:mcHRN}
    \langle\cHR\rangle_{\sN} &= \frac{%
      N_{\epsilon,L} \langle\cHR\rangle_{\epsilon,L} + N_{\epsilon,H}
      \langle\cHR\rangle_{\epsilon,H}}{%
      N_{\epsilon,L} + N_{\epsilon,H}} = F^{\epsilon}_{L}
    \langle\cHR\rangle_{\epsilon,L} +
    F^{\epsilon}_{H} \langle\cHR\rangle_{\epsilon,H} \\
    \label{eq:mcHRS}
    \langle\cHR\rangle_{\sS} &= \frac{%
      N_{\alpha,L} \langle\cHR\rangle_{\alpha,L} + N_{\alpha,H}
      \langle\cHR\rangle_{\alpha,H}}{%
      N_{\alpha,L} + N_{\alpha,H}} = F^{\alpha}_{L}
    \langle\cHR\rangle_{\alpha,L} + F^{\alpha}_{H}
    \langle\cHR\rangle_{\alpha,H} \\
    \label{eq:mcHRH}
    \langle\cHR\rangle_{H} &= \frac{%
      N_{\epsilon,H} \langle\cHR\rangle_{\epsilon,H} + N_{\alpha,H}
      \langle\cHR\rangle_{\alpha,H}}{%
      N_{\epsilon,H} + N_{\alpha,H}} = F^{H}_{\epsilon}
    \langle\cHR\rangle_{\epsilon,H} + F^{H}_{\alpha}
    \langle\cHR\rangle_{\alpha,H} \\
    \label{eq:mcHRL}
    \langle\cHR\rangle_{L} &= \frac{%
      N_{\epsilon,L} \langle\cHR\rangle_{\epsilon,L} + N_{\alpha,L}
      \langle\cHR\rangle_{\alpha,L}}{%
      N_{\epsilon,L} + N_{\alpha,L}} = F^{L}_{\epsilon}
    \langle\cHR\rangle_{\epsilon,L} + F^{L}_{\alpha}
    \langle\cHR\rangle_{\alpha,L}
  \end{align}
  Combining these we obtain the equations for the \Ha{} bias and mass
  step in terms of the partitioned data:
  \begin{align}
    \label{eq:Habias_def_general}
    \DcHR &= F^{\epsilon}_{L} \langle\cHR\rangle_{\epsilon,L} +
    F^{\epsilon}_{H} \langle\cHR\rangle_{\epsilon,H} - F^{\alpha}_{L}
    \langle\cHR\rangle_{\alpha,L} -
    F^{\alpha}_{H} \langle\cHR\rangle_{\alpha,H} \\
    \label{eq:mass_step_def_general}
    \DcHRMS &= F^{H}_{\epsilon} \langle\cHR\rangle_{\epsilon,H} +
    F^{H}_{\alpha} \langle\cHR\rangle_{\alpha,H} - F^{L}_{\epsilon}
    \langle\cHR\rangle_{\epsilon,L} - F^{L}_{\alpha}
    \langle\cHR\rangle_{\alpha,L}
  \end{align}
  With this development it is now possible to write the mass step
  directly in terms of the \Ha{} bias, as follows:
  \begin{equation}
    \label{annex:eq:ha-to-mass-step}
    \DcHRMS = \DcHR \times
    \frac{F^{H}_{\epsilon} \langle\cHR\rangle_{\epsilon,H} -
      F^{L}_{\epsilon} \langle\cHR\rangle_{\epsilon,L} +
      F^{H}_{\alpha} \langle\cHR\rangle_{\alpha,H} -
      F^{L}_{\alpha} \langle\cHR\rangle_{\alpha,L}}{%
      F^{\epsilon}_{H} \langle\cHR\rangle_{\epsilon,H} +
      F^{\epsilon}_{L} \langle\cHR\rangle_{\epsilon,L} -
      F^{\alpha}_{H} \langle\cHR\rangle_{\alpha,H} -
      F^{\alpha}_{L} \langle\cHR\rangle_{\alpha,L}}
  \end{equation}
  Note that in this expression we have simply rearranged terms in
  order to demonstrate that the same weighted mean magnitudes, just
  having different signs and fractions, appear in both the numerator
  and denominator.

  \subsection{Observational constraints}
  \label{sec:bias_math_obs}

  Eq.~\ref{annex:eq:ha-to-mass-step} is exact when the all of the
  quantities can be directly measured, as they are in our dataset.
  But, the variables can be generalized to any similar dataset.  In
  order to be exact, though, local environment observations would be
  required.  However, in the main text we presented two observational
  constraints that can be used to greatly simply
  Eq.~\ref{annex:eq:ha-to-mass-step}.  First, we have shown in
  Sect.~\ref{sec:mass-step} that the averaged \cHR{} of SNe~\sS{} does
  not depend on the masses of their hosts:
  \begin{equation}
    \label{eq:obs1}
    \langle\cHR\rangle_{\alpha,L} \simeq
    \langle\cHR\rangle_{\alpha,H} \simeq \langle\cHR\rangle_{\sS}.
  \end{equation}
  Second, we observed in Fig.~\ref{fig:mass-step} that SNe~\sS{} and
  \sN{} from low-mass hosts share the same mean magnitude (within
  error bars):
  \begin{equation}
    \label{eq:obs2}
    \langle\cHR\rangle_{\epsilon,L} \simeq
    \langle\cHR\rangle_{\alpha,L} \simeq \langle\cHR\rangle_{\sS}.
  \end{equation}
  As a consequence of these two observations, the right hand side of
  Eq.~\ref{eq:Habias_def_general} becomes:
  \begin{align}
    \label{eq:conseq:assumption_no_M2_low}
    F^{\epsilon}_{H} \langle\cHR\rangle_{\epsilon,H} +
    F^{\epsilon}_{L} \langle\cHR\rangle_{\epsilon,L} -
    \left(F^{\alpha}_{L} \langle\cHR\rangle_{\alpha,L} +
      F^{\alpha}_{H} \langle\cHR\rangle_{\alpha,H}\right) &\simeq
    F^{\epsilon}_{H} \langle\cHR\rangle_{\epsilon,H} +
    \left(1-F^{\epsilon}_{H}\right)\langle\cHR\rangle_{\sS} -
    \langle\cHR\rangle_{\sS} \nonumber\\
    &\simeq F^{\epsilon}_{H} \times
    \left(\langle\cHR\rangle_{\epsilon,H} -
      \langle\cHR\rangle_{\sS}\right).
  \end{align}
  Similarly, the right hand side of Eq.~\ref{eq:mass_step_def_general}
  becomes:
  \begin{align}
    \label{eq:mass_step_def_SNf}
    F^{H}_{\epsilon} \langle\cHR\rangle_{\epsilon,H} + F^{H}_{\alpha}
    \langle\cHR\rangle_{\alpha,H} - \left(F^{L}_{\epsilon}
      \langle\cHR\rangle_{\epsilon,L} + F^{L}_{\alpha}
      \langle\cHR\rangle_{\alpha,L}\right) &\simeq F^{H}_{\epsilon}
    \langle\cHR\rangle_{\epsilon,H} +
    \left(1-F_{\epsilon}^{H}\right)\langle\cHR\rangle_{\sS} -
    \langle\cHR\rangle_{\sS}\nonumber\\
    &\simeq F_{\epsilon}^{H} \times
    \left(\langle\cHR\rangle_{\epsilon,H} -
      \langle\cHR\rangle_{\sS}\right).
  \end{align}
  Finally, using the relation ${F^{H}_{\epsilon}}/{F^{\epsilon}_{H}} =
  (N_{\epsilon,H}/N_{H})\times(N_{\epsilon}/N_{\epsilon,H}) =
  (N_{\epsilon}/N)\times(N/N_{H}) = F_{\epsilon} /{F_{H}}$, where
  $F_{\epsilon}$ is the fraction of SNe~\sN{}, and $F_{H}$ the
  fraction of SNe~Ia from high-mass hosts,
  Eq.~\ref{annex:eq:ha-to-mass-step} can simply be expressed as:
  \begin{equation}
    \DcHRMS \simeq F_{\epsilon} \times \frac{\DcHR}{F_{H}}
  \end{equation}
  In the main text, the dependence of $F_{\epsilon}$ with redshift is
  denoted $\psi(z)$, and \DcHR{} is assumed to be constant.
  Furthermore, based on the dataset we complied in
  \cite{childress_host_2012}, $F_{H}$ appears to be constant within
  10\% up to $z\sim1$. Subject to these approximations, we conclude
  that the mass-step evolution is simply proportional to $\psi(z)$, as
  given in Eq.~\ref{eq:mass-step_z}.
\end{appendix}

\end{document}